\documentclass[10pt]{article}

\usepackage{graphicx}
\usepackage{amsmath}
\usepackage{amssymb}
\usepackage{amsthm}
\usepackage{color}
\usepackage{comment} 
\usepackage{url}
\usepackage[numbers]{natbib}

\newtheorem{thm}{Theorem}
\newtheorem{lem}{Lemma}
\newtheorem{prop}{Proposition}

\newtheorem{defn}{Definition}
\newtheorem{rem}{Remark}

\newcommand{\Def}[1]{\begin{defn}\,({\textsc #1})\,}
\newcommand{\EndDef}{\end{defn}\vspace{-0.0cm}}
\newcommand{\Rem}[1]{\begin{rem}\,({\textsc #1})\,}
\newcommand{\EndRem}{\end{rem}\vspace{-0.0cm}}

\definecolor{MyDarkBlue}{rgb}{0.2,0.2,0.6}
\newcommand{\ProofCore}{\begin{proof}}
\newcommand{\EndProofCore}{\end{proof}\color{black}}

\specialcomment{Proof}{\ProofCore}{\EndProofCore}

\newcommand{\eg}{e.g.,\,}
\newcommand{\ie}{i.e.,\,}

\newcommand{\SmallVJump}{\vspace{0.2cm}}

\newcommand{\OrdRel}{\preccurlyeq}

\newcommand{\Le}{\ell_{E}}
\newcommand{\Lv}{\ell_{V}}
\newcommand{\Lep}[1]{\ell_{E#1}}
\newcommand{\Lvp}[1]{\ell_{V#1}}

\newcommand{\OrdRelEG}{\preccurlyeq^e}

\newcommand{\Subgraph}{\subseteq}
\newcommand{\ISubgraph}{\subseteq_i}
\newcommand{\ESubgraph}{\subseteq_e}

\newcommand{\SV}{\Sigma_V}
\newcommand{\SE}{\Sigma_E}
\newcommand{\SubSV}{\OrdRel_{\SV}}
\newcommand{\SubSE}{\OrdRel_{\SE}}
\newcommand{\SizeSV}{s_{\SV}}
\newcommand{\SizeSE}{s_{\SE}}
\newcommand{\MV}{{\cal{M}}_V}
\newcommand{\MVMOM}{(\SV,\SubSV,\SizeSV)}
\newcommand{\ME}{{\cal{M}}_E}
\newcommand{\MEMOM}{(\SE,\SubSE,\SizeSE)}

\newcommand{\Model}[1]{{{\cal M}_{#1}}}

\newcommand{\MCSM}{MCS Model}
\newcommand{\MCSMS}{MCS Models}

\newcommand{\PS}{{\cal P}}

\newcommand{\Ev}{\varepsilon_V}
\newcommand{\Ee}{\varepsilon_E}

\newcommand{\Cn}{\kappa_n}
\newcommand{\Cnn}{\kappa_{2n}}
\newcommand{\GC}[2]{\kappa_{#1}^{#2}}

\newcommand{\Da}{d_{a}}
\newcommand{\Db}{d_{b}}
\newcommand{\Dc}{d_{c}}
\newcommand{\Dd}{d_{d}}

\newcommand{\MCSext}{Maximum Common Subelement Model}

\newcommand{\Subg}{\subseteq}
\newcommand{\ISubg}{\subseteq_i}

\newcommand{\CSub}{cs}
\newcommand{\CSubp}[1]{cs(#1)}
\newcommand{\CSubpp}[1]{cs(\{#1\})}

\newcommand{\Ox}{\!\OrdRel_{{\! {\kern+1pt}{\scalebox{0.6}{X}}{\kern+1pt}}}\!}

\newcommand{\Dged}{d_{\text{GED}}}

\newcommand{\DBS}{d_{B}}
\newcommand{\DW}{d_{W}}
\newcommand{\DFV}{d_{F}}

\newcommand{\AuthSep}{\hspace{0.5cm}} 

\newcommand{\FigureOne}{
\begin{figure}[p!]
\begin{center}
  \includegraphics[width=9cm]{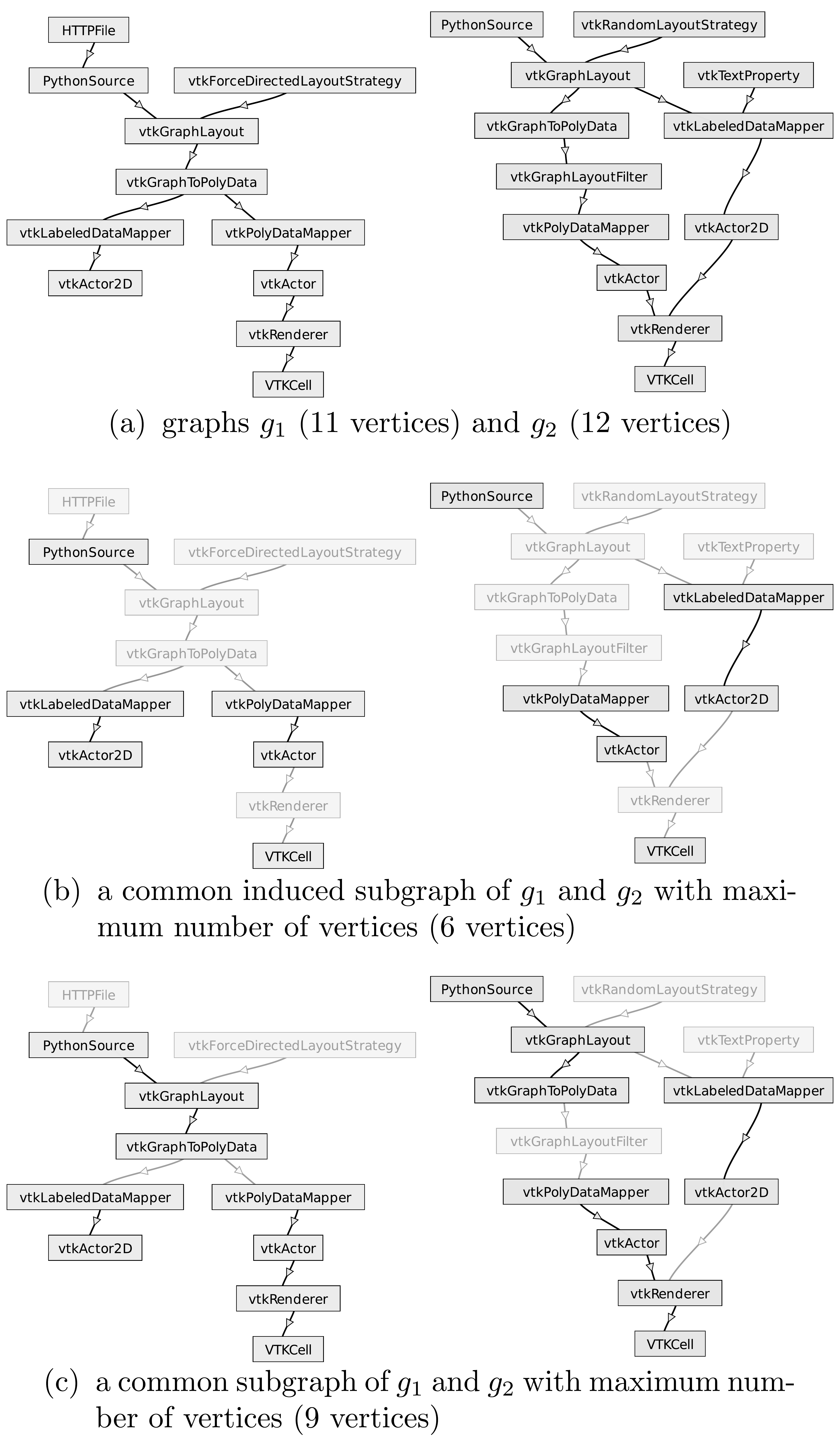}
\end{center}
\vspace{-0.5cm}
\caption{Difference between the maximum number of vertices of a common
  induced subgraph and of a common subgraph of graphs $g_1$ and $g_2$
  shown in (a). No common induced subgraph of $g_1$ and $g_2$ has more
  than 6 vertices (b), while there exist a common subgraph with 9
  vertices (c). (these graphs represent scientific workflow
  descriptions generated using \cite{vistrails} (2008))}
\label{fig:mcis-vs-mcs}
\end{figure}}

\newcommand{\FigureTwo}{
\begin{figure}[h!]
\begin{center}
  \includegraphics[width=8cm]{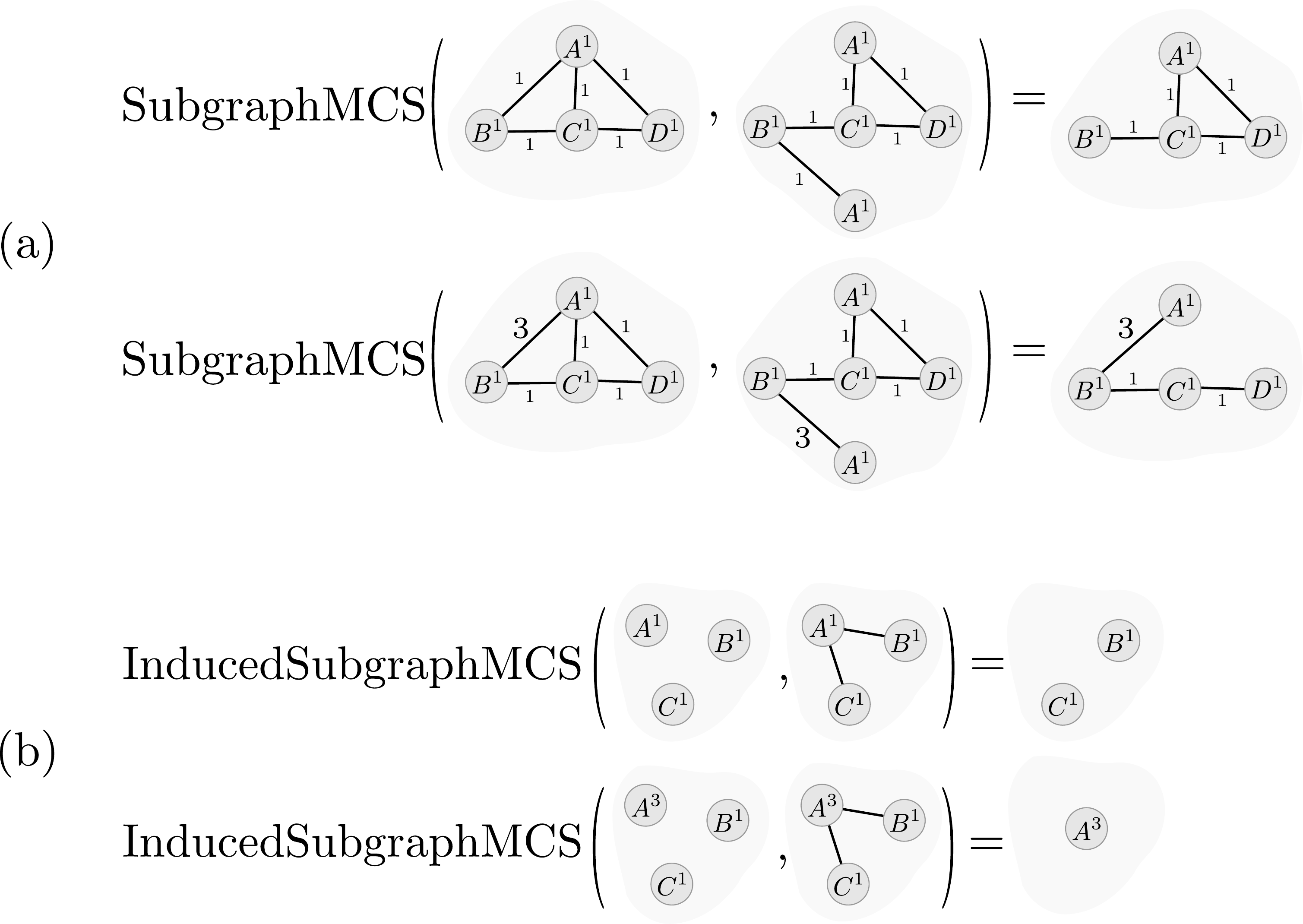}
\end{center}
\vspace{-0.5cm}
\caption{Effect of $\alpha$ (label weighting) on the notion of maximum common
  subelement of (a) S-\MCSM\ and (b) I-\MCSM. Vertex labels are
  letters superscripted with their $\alpha$ values. Edge labels match
  their $\alpha$ value in (a) and are ommited in (b), since they are
  not considered by $s_{GV\alpha}$. Note that depending on $\alpha$,
  the maximum common subelement changes.}
\label{fig:diff-mcs-notions}
\end{figure}}

\newcommand{\FigureThree}{
\begin{figure}[p!]
\begin{center}
  \includegraphics[width=0.8\textwidth]{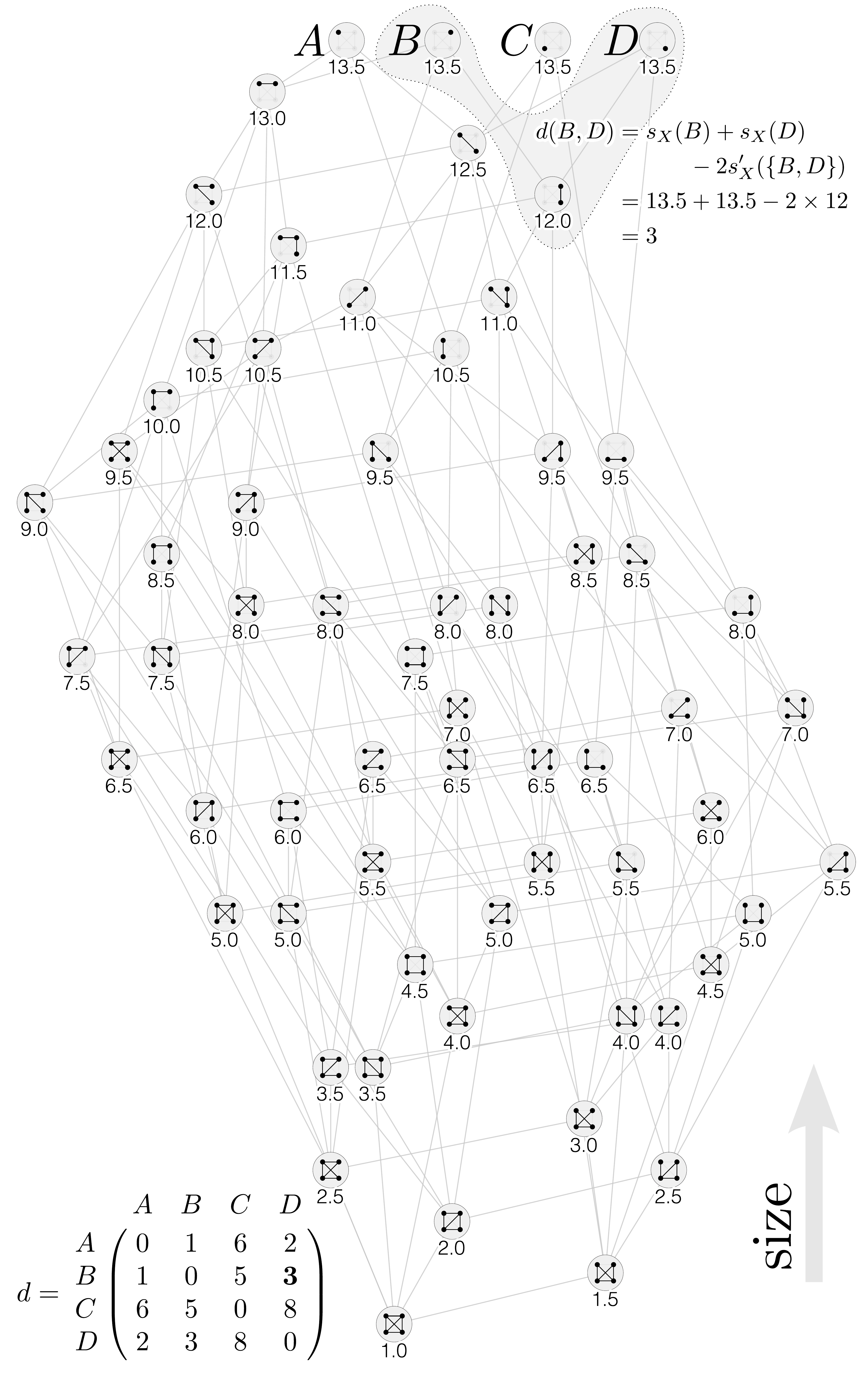}
\end{center}
\vspace{-0.5cm}
\caption{Metric $d$ on the set $\Sigma = \{A, B, C, D\}$ and
  \MCSM\ $\Model{X}=(X,\OrdRel_X,s_X)$ on a set $X$ where
  \hbox{$\Sigma \subseteq X$}.  Each element of $X$ is represented by
  a circle and $A, B, C, D$ are the top four circles. Two elements
  $x_1, x_2$ in $\Model{X}$ are related by $x_1 \! \OrdRel_X \! x_2$
  if there is an upward path from $x_1$ to $x_2$. The size function
  $s_X$ grows bottom-up and its values are shown below each
  corresponding element. Model $\Model{X}$ is related to $d$ by the
  fact that $d(\sigma_1, \sigma_2) = s_X(\sigma_1) + s_X(\sigma_2) - 2
  s'_X(\{ \sigma_1, \sigma_2\})$ for any $\sigma_1, \sigma_2$ in
  $\Sigma$. By Lemma~\ref{lem:dm2mom} for any other metric (on a
  finite set) there is a \MCSM\ satisfying the same properties as in
  this example.}
\label{fig:distmat2mcsm}
\end{figure}}

\title{Maximum Common Subelement Metrics and its Applications to Graphs}

\author{
Lauro Lins$^\ast$  \AuthSep Nivan Ferreira Jr.$^\dagger$ \AuthSep Claudio Silva$^\dagger$ \AuthSep Juliana Freire$^\dagger$ \\[0.1cm]
{\centering \small \em $\ast$ \!\!\! \{lauro.lins\}@gmail.com, \,\,\, $\dagger$ \!\!\! \{nivan.ferreira, juliana.freire, csilva\}@nyu.edu }
}

\begin{document}

\maketitle



\begin{abstract}
In this paper we characterize a mathematical model called {\em Maximum
  Common Subelement (MCS) Model} and prove the existence of four
different metrics on such model.
We generalize metrics on graphs previously proposed in the literature
and identify new ones by showing three different examples of
\MCSMS\ on graphs based on (1) subgraphs, (2) induced subgraphs and
(3) an extended notion of subgraphs.
This latter example can be used to model graphs with complex labels
(\eg graphs whose labels are other graphs), and hence to derive
metrics on them. Furthermore, we also use (3) to show that graph edit
distance, when a metric, is related to a maximum common subelement in
a corresponding \MCSM.

\end{abstract}



\section{Introduction}
\vspace{0.2cm}

Graphs are a natural model for a number of concepts in many different
domains such as
molecules in chemistry, interaction networks in social studies and
biochemistry, workflow descriptions in scientific computing, just to
name a few.
In each of these domains, when dealing with collections of such
objects, it is usually important to have a precise notion of
similarity/dissimilarity between them. 
An adequate and precise way to define similarity/dissimilarity
between graphs is by means of a {\em metric} on (the set of) graphs.

\citet{bunke1998gdm} showed that the function
$$ \DBS(g_1,g_2) = 1 - \frac{v_{12i}}{\max\{v_1, v_2\}}$$ is a metric
on the set of graphs when $g_1$ and $g_2$ are graphs with,
respectively, $v_1$ and $v_2$ vertices, and $v_{12i}$ is the maximum
number of vertices of a {\it common induced subgraph of $g_1$ and
  $g_2$}. Later, \citet{wallis2001graph} showed that, by rearranging the
same terms, the function $$ \DW(g_1,g_2) = 1 - \frac{v_{12i}}{v_1 +
  v_2 - v_{12i}}$$ is also a metric on the set of graphs.
We say that these two metrics are {\it based on induced subgraphs}
because the term $v_{12i}$ is related to a common induced subgraph of
input graphs $g_1$ and $g_2$.

An initial motivation for this work was to identify metrics on the set
of graphs {\it based on subgraphs} instead of induced subgraphs that
would be analogous $\DBS$ and $\DW$.
Note in Figure~\ref{fig:mcis-vs-mcs} that, for the same two graphs,
the largest number of vertices of a common subgraph and of a common
induced subgraph can be significatively different.
This observation leads to the fact that, depending on the application,
the graph similarity/dissimilarity notion is better modeled either by
a function based on subgraphs or by one based on induced subgraphs.
One application where a function based on subgraphs is a better fit is
reported by \cite{raymond2002mcs}. In their paper, they argue that a
common subgraph (not necessarily an induced one) that has the largest
number of edges is a better model for the similarity of chemical
graphs since, in their words, {\it ``it is the bonded interactions
  between atoms in a molecule that are the most responsible for its
  perceived activity''}.
In this application for chemical graphs, analogous versions of $\DBS$
and $\DW$ based on subgraphs would be more adequate.

One metric on the set of graphs based on subgraphs was shown by
\citet{fernandez2001graph}. Their function is equivalent to the
following definition:
$$\DFV(g_1, g_2) = (v_1 + e_1) + (v_2 + e_2) - 2 (v_{12s} +
e_{12s}),$$ where the new terms $e_1$ and $e_2$ are the number of
edges of $g_1$ and $g_2$, and $v_{12s}$ and $e_{12s}$ are the number
of vertices and edges of a common subgraph of $g_1$ and $g_2$ that
maximizes the sum of number of vertices and number of edges among all
subgraphs of $g_1$ and $g_2$.

In this paper we characterize a mathematical structure called {\em
  Maximum Common Subelement (MCS) Model}
(Section~\ref{sec:mcsmetrics}), that generalizes the one described by
\cite{deraedt2009ddm}, and show that four metrics are valid in such
model (Theorem~\ref{thm:metrics}), including general analogous
versions of the functions $\DBS$, $\DW$, and $\DFV$. We then show
three examples of \MCSMS\ on graphs. The first two examples are based
on the usual notions of subgraphs (Section~\ref{sec:gsmcs}) and
induced subgraphs (Section~\ref{sec:gismcs}), and the third example is
based on a notion of {\it extended} subgraphs
(Section~\ref{sec:gesmcs}).  We refer to these three MCS Model on
graphs as, respectively, {\it S-\MCSM}, {\it I-\MCSM}, and
{\it E-\MCSM}.  The importance of these \MCSMS\ on graphs is that
they enable us to reproduce previous metrics on graphs (\eg $\DBS$,
$\DW$, $\DFV$), extend them (weighting scheme), and derive new ones
(\eg analogous of $\DBS$ and $\DW$ based on subgraphs, the metrics
on the E-\MCSM).

One interesting aspect of the E-\MCSM\ is that the (vertex and edge)
labels of its graphs are elements of other \MCSMS. This permits an
E-\MCSM\ to describe {\it rich structured objects} (\eg graphs whose
labels are other graphs) and similarity models on them (\ie the
general \MCSM\ metrics are readly available for these rich structured
objects). In Section~\ref{sec:ged}, we use E-\MCSMS\ to show that for
any graph edit distance that is a metric on graphs, we can derive a
corresponding MCS Model where the edit distance of two graphs is
related to the size of a maximum common subelement of the two graphs
in this corresponding \MCSM.



\FigureOne

\section{Preliminaries}
\label{sec:preliminaries}
For the sake of completeness, in this section we state some standard
concepts that are fundamental for the rest of the paper.

\Def{{Metric, Metric Space}}
A {\it metric} $d$ on a set X is a function $d\!:\!X\!\times
\!X\!\rightarrow\![0,\infty)$ that, for any $x_1,x_2,x_3 \in X$, the
following conditions hold
\begin{align*}
\text{{(M1)} } & d(x_1,x_1) = 0; \\
\text{{(M2)} } & d(x_1,x_2) = d(x_2,x_1); \\
\text{{(M3)} } & d(x_1,x_3) \leq d(x_1,x_2) + d(x_2,x_3); \\
\text{{(M4)} } & \text{if } d(x_1,x_2) = 0 \text{ then } x_1 = x_2.
\end{align*}
In this case, the pair $(X,d)$ is called a {\it metric space}. If $X$
is finite then we also refer to $(X,d)$ as a {\it finite metric
  space}.  \EndDef

\Def{{Partial Order}} Let $\,\OrdRel\,$ be a relation on a set $X$,
    {\it i.e.}  $\,\OrdRel\,$ is a subset of $X \times X$. 
    We use the notation $x_1 \OrdRel x_2$ to mean $(x_1,x_2)$ 
    is an element of $\OrdRel$. We say $\OrdRel$
    is a {\it partial order on $X$} if the following conditions hold:
\begin{align*}
  \text{{(R1)} } & x \OrdRel x \text{ (reflexivity)} \\
  \text{{(R2)} } & x_1 \OrdRel x_2 \text{ and } x_2 \OrdRel x_3 \text{ then } x_1 \OrdRel x_3 \text{ (transitivity)} \\
  \text{{(R3)} } & x_1 \OrdRel x_2 \text{ and } x_2 \OrdRel x_1 \text{ then } x_1 = x_2 \text{ (antisymmetry)}
\end{align*}
\EndDef

Furthermore, we use the notations $|A|$, $\PS(A)$, and $[A]^k$ to
mean, respectively, the number of elements in set $A$, the {\it power
  set of $A$}, and the set of all sets containin $k \geq 1$ elements
of $A$.

\section{Maximum Common Subelement (MCS) Model}
\label{sec:mcs}

In general, a natural model to the similarity of two objects is
given by a number reflecting {\it how much} do the two objects {\it
  overlap}.
The {\it Maximum Common Subelement (MCS) Model} is a precise way of
encoding this idea of similarity, framed in a general
language that can fit many different scenarios (our focus application
in the following sections are graphs).
A MCS Model is composed of three parts. The first part is a {\it set}
also called the {\it domain} of the model. The second part, used to to
make the informal notion of {\it overlap} precise, is a {\it partial
  order} on the set or domain of the model. The third and last part,
used to quantify {\it how much is an overlap}, is a {\it size
  function} which assigns a {\it size value} for each element of the
domain. The following definitions fix some notation before we formally
define a \MCSM.



\Def{{Subelement, Superelement, Common Subelements}} Let $\OrdRel$ be
a partial order on a set $X$.  For $x_1, x_2 \in X$, if $x_1 \OrdRel
x_2$, we say that $x_1$ is a {\it subelement} of $x_2$ and that $x_2$
is a {\it superelement} of $x_1$. We define the function {\it common
  subelements}, denoted by $\CSub$, as
  \begin{align*}\CSubp{X'} = \{\,\, x \in X: x \OrdRel
    y, \forall y \in X'\}, \,\, \text{ for } X' \subseteq X.
  \end{align*}
\EndDef

\Def{{Size Function}} Let $\OrdRel$ be a partial order on $X$.  We say
a function $s:X \rightarrow [0,\infty)$ is a {\em size function} on
  $(X,\OrdRel)$ if, for $x_1, x_2 \in X$, the following conditions
  hold
\begin{align*}
\text{{(S1)} } & \text{if } x_1 \OrdRel x_2 \text{ then } s(x_1) \leq s(x_2); \\
\text{{(S2)} } & \text{if } x_1 \OrdRel x_2 \text{ and } s(x_1) = s(x_2) \text{ then } x_1 = x_2. 
\end{align*}
\EndDef

The size function conditions (S1) and (S2) formalizes the idea that a
subelement must have either a smaller size ({\it proper} subelement),
or have the same size and be the same element (a {\it non-proper}
subelement). Now we are ready to define a \MCSM.

\Def{{Maximum Common Subelement Model}} \label{def:sos_model}A {\it
  Maximum Common Subelement (MCS) Model on a set $X$} is a
triple $$(X,\OrdRel,s),$$ where $\OrdRel$ is a partial order on $X$,
and $s$ is a size function on $(X,\OrdRel)$ such that
\vspace{-0.2cm}
\begin{align*}
\text{{(A1)}} \,\, & \text{Given } x_1, x_2 \in X, \,\,  \CSubpp{x_1,x_2} \neq \emptyset \text{ and } \\
   & \hspace{0.5cm}  \{s(x) \, | \, x \in \CSubpp{x_1,x_2}\} \text{ has a maximum}; \\ 
\text{{(A2)}} \,\, & \text{Given } x_1,x_2,x \in
X \text{ and } x_1, x_2 \OrdRel x \text{ there exists } x_{12} \in
\CSubpp{x_1,x_2} \\ & \hspace{0.5cm} \text{such that } s(x) \geq s(x_1) + s(x_2) -
s(x_{12}).
\end{align*}
\EndDef 
\vspace{0.3cm}

Condition (A1) on a \MCSM\ states that any two elements (not
necessarily distinct) have at least one common subelement, and, among
all common subelements, there is at least one (could be more than one)
whose size is the largest possible. Condition (A2) is rooted on the
idea that a superelement of any two elements must, some how, contain
these two elements simultaneously, in other words, it contain a kind
of {\it union} of these two elements. Imagine two finite sets $S_1$
and $S_2$, intuitively we expect that the number of elements of any
superset $S$ of sets $S_1$ and $S_2$ to have at least as much elements
as their union: $|S| \geq |S_1 \cup S_2| = |S_1| + |S_2| - |S_1 \cap
S_2|$, but never fewer elements than that.

\newcommand{\DRM}{DR Model} 

The \MCSM\ is a generalization of the model proposed by
\cite{deraedt2009ddm}, referred here as the {\it \DRM}. The motivation
to define the \DRM\ in their paper was the same we had to define the
MCS Model here: a template to fit applied situations into, and derive
metrics. A terminology difference between the \DRM\ and the \MCSM\
is that the terms {\it pattern, generalization, specialization} in the
former becomes, respectivelly, {\it element, subelement, superelement}
in the latter. A more important difference is that, in our
terminology, while the \DRM\ requires every two elements to have at
least one subelement and one superelement, the MCS Model only requires
the subelement to exist. Once the terminology between the two models
is aligned, it is straightfoward to prove that MCS Model is in fact a
generalization of the \DRM\ (\eg {\it diamond inequality} there is
equivalent to condition (A2)). Thus, all the examples given in that
paper, namely weighted sets, strings and trees (with appropriate
partial order relations and size functions) are also examples of MCS
Models.  \cite{deraedt2009ddm} proved one metric function to be valid
in any \DRM. In this paper (Theorem~\ref{thm:metrics}) we extend this
list to four metric functions to be valid in an even more general
model: the MCS Model.

When presenting examples of \MCSMS\ in the following sections,
instead of showing that property (A1) is valid, we show that the
following more restrictive property (A1') is valid:
\begin{align*}
\text{{(A1')} \,\, } & 0 < |\CSubpp{x_1,x_2}| < \infty;
\end{align*}
Clearly, (A1') implies (A1), since the number of subelements of any
two elements is finite.

Before going into some properties and the metrics of \MCSMS\ 
we set more terminology 

\Def{{Auxiliar Functions}}
\label{def:maxcommonsubelementssize}
If \hbox{$(X, \OrdRel, s)$} is a MCS Model and $X'$ is a subset
of $X$ then the {\it maximum common subelements size function},
denoted by $s'$, is defined by
$$s'(X') = \max\{\,\, s(x): x \in cs(X') \,\, \},$$ and the {\it maximum
  common subelements function}, denoted by $mcs$, is defined by
$$ mcs(X') = \{\,x : s(x) = s'(X'), x \in cs(X') \, \}.$$
 \EndDef

Note that, in general, $s'$ and $mcs$ might not be well defined (\eg
common subelements of three elements might be empty). By (A1), these
functions are well defined when when $|X'| \leq 2$. In the rest of the
paper we should use these functions only when they are well defined.


\vspace{0.3cm}
\phantom{a}

\subsection{Some Properties of MCS Models}
\label{sec:productsosmodel}
\vspace{0.3cm}


\begin{prop}[{\sc Uniqueness of minimum size element}]
  \label{prop:uniqzero}
  Let \hbox{$(X,\OrdRel,s)$} be a MCS Model and let
  $x_0 \in X$ be such that $s(x_0) = min(\{s(x)|x \in X\})$. Then, the following statements
  are true:
  \begin{itemize}
    \item[(a)] if $x \in X$ is such that $s(x) = s(x_0)$, then $x=x_0$.
    \item[(b)] The element $x_0$ is a global subelement, {\it i.e.}, $x_0 \in cs(X)$.
  \end{itemize}
\end{prop}
\begin{Proof} 
  {(a)} Let $y \in cs(\{x_0,x\})$. It exists by $(A1)$. By (S1), we
  conclude that $s(y) = s(x_0) = s(x)$ and using (S2), we conclude
  that $x = y = x_0$. {(b)} Let $y \in X$ and, again, let $z \in
  cs(\{x_0,y\})$.  By (S1), we conclude that $s(z) = s(x_0)$, and
  using (S2), we have that $x_0 = z \Rightarrow x_0 \OrdRel y$.
\end{Proof}

\FigureThree

The following lemma provides a way to derive a MCS Model from a finite
metric space. The interesting relation between this metric space and
its derived MCS Model is that the metric is somehow preserved in the
structure of the MCS Model. Figure~\ref{fig:distmat2mcsm} presents a
finite metric space and a visual illustration of its derived MCS
Model. We use this lemma in Section~\ref{sec:ged} to built a relation
between Graph Edit Distance and MCS Models.


\begin{lem}[Metric Space to MCS Model]
\label{lem:dm2mom}
Let $\Sigma$ be a finite set and the function \hbox{$d:\Sigma \times \Sigma
  \rightarrow [0,\infty)$} be a metric on $\Sigma$. In this case,
  there is a \MCSM\ $$\Model{X} = (X,\Ox,s_X)$$ where $\Sigma
  \subseteq X$ and, for $\sigma_1,\sigma_2 \in \Sigma$,
\begin{equation}
\label{eq:dm2mom}
 d(\sigma_1, \sigma_2) = s_X(\sigma_1) + 
                  s_X(\sigma_2) - 2 s'_X(\{ \sigma_1, \sigma_2\}).
\end{equation}
\end{lem}
\newcommand{\EConn}{Z}
\begin{Proof} 
Let $n = |\Sigma|$ and $K_n = (\Sigma,[\Sigma]^2)$ be a
complete (unlabeled simple) graph. Assume the natural interpretation:
in $K_n$, an edge $\{\sigma_1, \sigma_2\} \in [\Sigma]^2$ has endpoints
  $\sigma_1, \sigma_2 \in \Sigma$. Furthermore, let $\EConn$ be all non
  empty subsets of edges in $K_n$ that induces a connected subgraph of
  $K_n$. We are now able to define the elements of our $\Model{X}$:
  the set $X$, the order relation $\Ox$, and the size function $s_X$.
  First, the elements of $X$ are the vertices of $K_n$ plus 
  every subset of edges in $K_n$ that induces a connected subgraph of $K_n$:
  $$X = \Sigma \cup \EConn.$$
  For $x_1, x_2 \in X$, let $x_1\Ox x_2$ if
  \begin{align*}
    (O1) \,\, & x_1 = x_2, \\
    (O2) \,\, & x_1 = E \subseteq [\Sigma]^2, x_2 = \sigma \in \Sigma, \\
       & \hspace{0.5cm} \text{ and vertex $\sigma$ 
      is an endpoint of some edge in $E$.} \\
    (O3) \,\,& x_1, x_2 \subseteq [\Sigma]^2 \text{ and } x_1 \supseteq x_2.
  \end{align*}
  Define $R$ to be
  $$R \, = \, \theta \,+ \,\frac{1}{2}{\sum_{\{\sigma_1,\sigma_2\}
      \atop \in [\Sigma]^2} {d(\sigma_1, \sigma_2)}}, $$ for some
  $\theta > 0$. For $x \in X$ define
  $$s_X(x) = \begin{cases}
    R,& \text{if } x \in \Sigma, \\
    {\displaystyle R - \frac{1}{2}\sum_{\{\sigma_1,\sigma_2\} \atop \in x}{d(\sigma_1,\sigma_2)},}& \text{for } x \subseteq [\Sigma]^2.,
  \end{cases}$$
  Note that, with this definition of $s_X$, $\theta$ is the size
  of $[\Sigma]^2$ (which is an element in $\EConn$ and is the
  smallest element in $\Model{X}$). We now prove that $\Model{X} = (X,
  \Ox, s_X)$ is a \MCSM:
\begin{itemize}
\item ({\textsc{$\Ox$ is a partial order}}) 
  \begin{itemize}
  \item ({\textsc{Reflexive}}) By (O1), $\Ox$ is reflexive.
  \item ({\textsc{Transitive}}) Assume $$\text{(H1)} \,\, x_1 \Ox x_2
    \,\,\, \text{ and \,\, (H2)} \,\, x_2 \Ox x_3.$$ If $x_1, x_2, x_3
    \in \EConn$, then, by $(O3)$, $x_3 \supseteq x_2 \supseteq x_1$,
    therefore, $x_3 \supseteq x_1$ and, again by (O3), $x_1 \OrdRel
    x_3$. Note that, if $x \in \Sigma$, then it is maximal on
    $\Ox$. Therefore, for $1 \leq i < j \leq 3$, if $x_i \in \Sigma$,
    then $x_j = x_i$.  If $x_1 \in \Sigma$ then $x_1 = x_2 = x_3$ and,
    by (O1), $x_1 \OrdRel x_3$. If $x_1 \in \EConn$ and $x_2 \in
    \Sigma$, then $x_2 = x_3$ and (H1) is equivalent to $x_1 \Ox x_3$.
    If $x_1,x_2 \in \EConn$ and $x_3 \in \Sigma$, then, by (H1), $x_1
    \supseteq x_2$ and, by (H2), $x_1$ is an endpoint of some edge $e$
    in $x_2$. As edge $e$ is also en edge in $x_1$ we can conclude
    $x_1 \Ox x_3$.
  \item ({\textsc{Antisymmetric}}) Assume
    $$\text{(H3)} \,\, x_1 \Ox x_2 \,\,\, \text{ and \,\, (H4)} \,\,
    x_2 \Ox x_1.$$ If $x_1 \in \Sigma$, then, by (H3), we must have
    $x_1 = x_2$. If $x_1,x_2 \in \EConn$, then  (H3) and (H4) 
    means $x_1 \supseteq x_1$ and $x_2 \supseteq x_2$ therefore $x_1 = x_2$.
  \end{itemize}
\item ({\textsc{$s_X$ is a size function}})
  First, by the definitions of $R$ and $s_X$ it is easy to
  check that $s_X(x) \geq 0$ for all $x \in X$.
  If $x_1 \Ox x_2$ then three cases can occur: 
$$\text{(C1) } x_1, x_2 \in Z, \,\,\, \text{(C2) } x_1 \in Z, x_2 \in
  \Sigma, \,\,\, \text{(C3) } x_1,x_2 \subseteq \Sigma.$$
  \begin{itemize}
  \item (S1) We need to show that:
    $$ \text{if } x_1 \Ox x_2 \text{ then } s(x_1) \leq s(x_2).$$ If
    (C1) occurs then $x_1 \supseteq x_1$ and, by definition, the
    expression for $s_X(x_1)$ will subtract from $R$ at least the same
    edge terms $d(\sigma_1, \sigma_2)/2$ as the expression for
    $s_X(x_1)$, therefore, $s_X(x_1) \leq s_X(x_2)$. If case (2)
    occurs, $x_1$ has at least one edge $e$ and the expression for
    $s_X(x_1)$ subtracts at least one positive value from $R$ (\eg
    the value relative to $e$). Since $x_2 = R$, then $s(x_1)
    \leq s(x_2)$. If case (3) occurs, then $x_1 = x_2$, therefore
    $s(x_1) \leq s(x_2)$.
  \item(S2) We need to show that:
    $$ \text{if } x_1 \OrdRel x_2 \text{ and } s(x_1) = s(x_2) \text{ then }
    x_1 = x_2. 
    $$ Assume $x_1 \Ox x_2$ and $s(x_1) = s(x_2)$.  If (C1) occurs,
    then $x_1 \supseteq x_2$ and, by definition, the expression for
    $s_X(x_1)$ subtracts from $R$ at least the same edges as in
    $s_X(x_2)$. If we assume $x_1 \neq x_2$ then $s_X(x_1)$ would
    subtract at least one more positive term from $R$ and therefore
    $s_X(x_1) < s_X(x_2)$, which contradicts the assumption $s_X(x_1)
    = s_X(x_2)$. Case (C2) cannot occur, because $x_1 \Ox x_2$, $x_1
    \in \EConn$, and $x_2 \in \Sigma$ would imply $s_X(x_1) <
    s_X(x_2)$ since $s(x_2) = R$ and $s(x_1)$ definition subtracts at
    least one positive term from $R$. If (C3) occurs, then
    necessarily $x_1 = x_2$.
  \end{itemize}
\item (A1) We are going that (A1) holds, by showing that (A1') also holds. 
  In order to do so, let $x_1,x_2 \in X$. By definition, the element 
  $x = [\Sigma]^2 \in X$ is a subelement of all other objects in $X$, 
  therefore $|cs(\{x_1,x_2\})| \geq 1$, for all $x_1, x_2 \in X$. 
  As $X$ is finite, then $|cs(\{x_1,x_2\})|$ must be finite.
\item (A2) We have to show that
\begin{align*}
  & \text{Given } x_1,x_2,x \in X \text{ and } x_1, x_2 \OrdRel x \\
  & \text{there exists } x_{12} \in cs(\{x_1,x_2\}) \\
  & \text{such that } s(x) \geq s(x_1) + s(x_2) - s(x_{12}).
  \end{align*}
  Before going into this axiom, we first note that if $x_1, x_2 \in
  \EConn$, $x \in X$ and $x_1, x_2 \Ox x$, then $x_1 \cup x_2$ is also
  an element of $\EConn$.  To see this fact, let $\sigma \in \Sigma$
  be equal to $x$, $x \in \Sigma$, or an endpoint of one edge in $x$
  if $x\in Z$.  Suppose $\sigma_a$ is a vertex in $x_1$ and $\sigma_b$
  is a vertex in $x_2$ (both are in $x_1 \cup x_2$). There must be a
  path from $\sigma_a$ to $\sigma$ in $x_1$ (which is contained in
  $x_1 \cup x_2$) and there must be a path from $\sigma_b$ to $\sigma$
  in $x_2$ (which is contained in $x_1 \cup x_2$).  By joining these
  paths we have a path from $\sigma_a$ to $\sigma_b$ in $x_1 \cup
  x_2$. Therefore, $x_1 \cup x_2 \in \EConn$.  

  We will split this into three cases: (1) $x_1 \in \Sigma$; (2) $x
  \in \EConn$; (3) $x \in \Sigma$ and $x_1, x_2 \in \EConn$.
  By the symmetric roles that $x_1$ and $x_2$ take in this axiom,
  these cases are enough to cover all possibilities.
  \begin{itemize}
  \item Case (1): As $x_1 \in \Sigma$, then we must have $x = x_1$ and
    making $x_{12} = x_2$ we have the axiom, since 
\begin{align*}
  s_X(x) \geq s_X(x) & = s_X(x_1) \\
  & = s_X(x_1) + s_X(x_2) + s_X(x_2) \\ 
  & = s_X(x_1) + s_X(x_2) + s_X(x_{12}).
\end{align*}
  \item Case (2): As $x \in \EConn$, then $x_1 \supseteq x$, $x_2
    \supseteq x$. Note that $x_1 \cap x_2 \supseteq x$ and $x$ has at
    least one edge since it is an element in $\EConn$. Making $x_{12}
    = x_1 \cup x_2$ we have the axiom since:
    \begin{align*}
      s_X(x) & s_X(x_1 \cap x_2)\\ 
      & = s_X(x_1) + s_X(x_2) + s_X(x_{12}).
    \end{align*}
    Note that $x_1 \cap x_2$ might not be a member of $\EConn$, but,
    since the formula for $s_X$ is well defined any subset of
    $[\Sigma]^2$, we used it.
  \item Case (3): Note that the inequation $$s_X(x) \geq s_X(x_1 \cap
    x_2)$$ is also true in this case. If $x_1 \cap x_2$ is the empty
    set we have that the only element in common between the graphs
    induced by $x_1$ and $x_2$ is the single vertex $x$. We can use
    the same development as in the previous case to estbilish the
    axiom in this case.
  \end{itemize}
\end{itemize}
It now remains to show that Equation~\ref{eq:dm2mom} is valid.  The
case where $\sigma_1 = \sigma_2$ is trivially true. Suppose $\sigma_1
\neq \sigma_2$. By the definition of $\Ox$ we know that $\{\{\sigma_1,
\sigma_2\}\} \Ox \sigma_1$ and $\{\{\sigma_1, \sigma_2\}\} \Ox
\sigma_2$. Furthermore, by the definition of $s_X$,
\begin{align*}
& s_X(\{\{\sigma_1,\sigma_2\}\}) = R - \frac{1}{2} d(\sigma_1, \sigma_2) \\
\equiv \,\, & 2 s_X(\{\{\sigma_1,\sigma_2\}\}) = 2 (R ) - d(\sigma_1, \sigma_2) \\
\equiv \,\, & d(\sigma_1, \sigma_2) = 2 (R  - 2 s_X(\{\{\sigma_1,\sigma_2\}\}) \\
\equiv \,\, & d(\sigma_1, \sigma_2) = s_X(x_1) + x_X(x_2) - 2 s_X(\{\{\sigma_1,\sigma_2\}\})
\end{align*}
which is in the form of Equation~\ref{eq:dm2mom}. If
$\{\{\sigma_1,\sigma_2\}\} \in mcs(\{\sigma_1, \sigma_2\})$ then we
have the result. Let's show that this is indeed true. Let $x$ be a
memeber of $mcs(\{\sigma_1, \sigma_2\})$. It then satisfies: $x \Ox
\sigma_1, \sigma_2$. It also must be in $\EConn$ since $\sigma_1 \neq
\sigma_2$. There must be a path
$\sigma_1,\beta_1,\beta_2,\ldots,\beta_k,\sigma_2$ in $x$ otherwise $x
\Ox \sigma_1$ and $x \Ox \sigma_2$ would not be true. Actually $x$
must induce a path from $\sigma_1$ to $\sigma_2$ otherwise we could
remove the extra (non-path) edges and still get a set of edges 
inducing a connected subgraph and with a larger $s_X$. Furthermore,
for paths of the form $\sigma_1,\beta_1,\beta_2,\ldots,\beta_k,\sigma_2$,
replacing edges $\{\sigma_1, \beta_1\}$ and $\{\beta_1, \beta_2\}$ by
 $\{\sigma_1, \beta_2\}$ we still have a path, and, by the fact that
$d$ is a metric, we have not increased the size $s_X$ of our $x$.
This way we can erase all intermediate graphs and get that 
the graph that induces the path $\sigma_1, \sigma_2$ must be a 
member of $mcs(\{\sigma_1, \sigma_2\})$. With this, the
result is estabilished. 
\end{Proof}



Lemma \ref{lem:aux} is a technical property used in
Section~\ref{sec:mcsmetrics} in the proof of
Theorem~\ref{thm:metrics}.

\begin{lem}
\label{lem:aux}
Let $(X,\OrdRel,s)$ be a \MCSM. The inequality 
\begin{equation}
\label{lem:aux} s'(\{x_1,x_2\}) + s'(\{x_2,x_3\}) \, \leq \, s(x_2) +
  s'(\{x_1,x_3\}).
\end{equation}
holds for all $x_1, x_2, x_3 \in X.$
\end{lem}
\begin{Proof} 
Let $x_{12} \in mcs(\{x_1,x_2\})$ and $x_{23} \in
mcs(\{x_2,x_3\})$. As $x_{12}, x_{23} \OrdRel x_2$, we can use axiom
(A2) to conclude that there exists $x_{123} \in cs(\{x_{12},x_{23}\})$
such that $s(x_2) \geq s(x_{12}) + s(x_{23}) - s(x_{123})$. We then
can write
\begin{align*}
s'(\{x_1,x_2\}) + s'(\{x_2,x_3\}) & = s(x_{12}) +
s(x_{23})  \\ & \leq s(x_2) + s(x_{123}) \leq s(x_2) + s'(\{x_1,x_3\}).
\end{align*}
\end{Proof}

\vspace{0.3cm}
\subsection{Metrics on MCS Models}
\label{sec:mcsmetrics}
\vspace{0.3cm}

The main result in this section is the Theorem \ref{thm:metrics} which
states that four different functions are indeed metrics on \MCSMS.

\begin{thm}[{\sc Metrics on MCS Models}] \label{thm:metrics}
Let $\Model{} = (X,\OrdRel,s)$ be a \MCSM\ on $X$ and let
$\Da, \Db, \Dc,\Dd$ be
\begin{align}  
\label{eq:symmdiffformula} 
\Da(x_1,x_2) &=  s(x_1) + s(x_2) - 2s'(\{x_1, x_2\}), \\[0.2cm]
\label{eq:maxminuscommonformula} 
\Db(x_1,x_2) &=  \max\{s(x_1),s(x_2)\} - s'(\{x_1, x_2\}), \\[0.2cm]
\label{eq:dc}
\Dc(x_1,x_2) &=  
     \begin{cases}
       \,\,0 \,, \,\,\text{ if } s(x_1) = s(x_2) = 0
       &  \\[0.4cm]
       \,\,\displaystyle 1 - \frac{s'(\{x_1, x_2\})}{\max\{s(x_1),s(x_2)\}} \,\,, \textrm{ otherwise. }
       & 
     \end{cases}\hspace{-1cm} \\[0.3cm] 
\label{eq:normunionformula}
\Dd(x_1,x_2) &=  
     \begin{cases}
       \,\,0 \,, \,\,\textrm{ if } s(x_1) = s(x_2) = 0 & \\[0.4cm] 
       \,\,\displaystyle 1 - \frac{s'(\{x_1, x_2\})}{s(x_1) + s(x_2) - s'(\{x_1,x_2\})} \,,
       \textrm{otherwise.}
       & 
     \end{cases} \hspace{-1.3cm}
\end{align}
Then, all of them are metrics on $X$.
\end{thm}
\begin{Proof} 
Since $s'(\{x_1,x_1\}) = s(x_1)$, it
is easy to check that (M1) is true for all formulas. Furthermore, as
$s'(\{x_1,x_2\}) =s'(\{x_2,x_1\})$, it is also easy to see that (M2)
is true for all formulas. Note that for any \MCSext\ we have
\begin{align}
\label{eq:m4aux}
x_1 \neq x_2 \,\, \Rightarrow \,\, s'(\{x_1,x_2\}) & < \max(\{s(x_1),s(x_2)\}) \\
&\leq s(x_1) + s(x_2) - s'(\{x_1,x_2\}). \nonumber
\end{align}
It is now easy to see that (M4) is true in all formulas (except for
the case $s(x_1) = s(x_2) = 0$ on $\Dc$ and $\Dd$) by using its
contrapositive form $$\text{if } x_1 \neq x_2 \text{ then } d(x_1,x_2)
\neq 0.$$ Assume $x_1 \neq x_2$ and check that, using
Equation~\ref{eq:m4aux} above, each of the four distance formulas will
result in a positive number. The case when $s(x_1) = s(x_2) = 0$ on
the formulas $\Dc$ and $\Dd$ is also true because, by
Proposition~\ref{prop:uniqzero}, there can be only one element with size
zero in a \MCSM. The proof of (M3) will be given separately for each
formula. For all the following proofs let $x_1,x_2,x_3 \in X$. 
\begin{itemize}
\item {(M3) \textsc{is valid for $\Da$:}} \,\, Using the Lemma \ref{lem:aux},
  we can write
\begin{align*}
       & 0 \, \leq \, s(x_2)  + s'(\{x_1,x_3\}) - s'(\{x_1,x_2\}) - s'(\{x_2,x_3\})  \\
\Rightarrow \,\,\, & 0 \leq 2s(x_2) + 2s'(\{x_1,x_3\}) - 2s'(\{x_1,x_2\})- 2s'(\{x_2,x_3\}) \\
\Rightarrow \,\,\, & s(x_1) + s(x_3) - 2s'(x_1,x_3) \leq s(x_1) + s(x_2) \\
       & \hspace{0.5cm}- s'(x_1,x_2) + s(x_2) + s(x_3)-2s'(x_2,x_3) \\
\Rightarrow \,\,\, & \Da(x_1,x_3) \leq \Da(x_1,x_2) + \Da(x_2,x_3).
\end{align*}
which proves (M3) for $\Da$.

\newcommand{\Sep}{\hspace{0.7cm}}
\item {(M3) \textsc{is valid for $\Db$:}} \,\, We split this proof in
three cases. These are the only cases need to be considered, since the 
role played by $x_1$ and $x_3$ in (M3) are symmetric.
\begin{itemize}
\item (Case 1) If $s(x_2)\leq s(x_1)\leq s(x_3)$ We can write:
\begin{align*}
  \text{Lemma \ref{lem:aux}} \Rightarrow \,\,\, & 
    0 \leq s(x_1) + s'(\{x_1,x_3\}) - s'(\{x_1,x_2\})- s'(\{x_2,x_3\}) \\
         \Rightarrow \,\,\, & s(x_3) - s'(\{x_1,x_3\}) \leq s(x_1) \\
  & \Sep - s'(\{x_1,x_2\}) + s(x_3)- s'(\{x_2,x_3\}) \\
  \Leftrightarrow \,\,\, & \max(\{s(x_1),s(x_3)\}) - s'(\{x_1,x_3\}) \leq \\
  & \Sep \max({\{s(x_1),s(x_2)\}}) - s'(\{x_1,x_2\}) \\
  & \Sep + \max({\{s(x_2),s(x_3)\}})- s'(\{x_2,x_3\}) \\
  \Rightarrow \,\,\, & \Db(x_1,x_3) \leq \Db(x_1,x_2) + \Db(x_2,x_3).
\end{align*}
\item (Case 2) If $s(x_1)\leq s(x_2)\leq s(x_3)$, adding $s(x_3)$ to
  both sides of (\ref{lem:aux}) we have
\begin{align*}
  \,\,\, & s(x_3) - s'(\{x_1,x_3\}) \leq s(x_2) \\
  & \Sep - s'(\{x_1,x_2\}) + s(x_3)- s'(\{x_2,x_3\}) \\
  \Leftrightarrow \,\,\, & \max(\{s(x_1),s(x_3)\}) - s'(\{x_1,x_3\}) \leq \\
  & \Sep \max({\{s(x_1),s(x_2)\}}) - s'(\{x_1,x_2\}) \\
  & \Sep + \max({\{s(x_2),s(x_3)\}})- s'(\{x_2,x_3\}) \\
  \Rightarrow \,\,\, & \Db(x_1,x_3) \leq \Db(x_1,x_2) + \Db(x_2,x_3).
\end{align*}
\item (Case 3) If $s(x_1)\leq s(x_3)\leq s(x_2)$, adding $s(x_3)$ to
  the left hand side and $s(x_2)$ to the right hand side of
  (\ref{lem:aux}) we have
\begin{align*}
  \,\,\, & s(x_3) - s'(\{x_1,x_3\}) \leq s(x_2) \\
  & \Sep - s'(\{x_1,x_2\}) + s(x_2)- s'(\{x_2,x_3\}) \\
  \Leftrightarrow \,\,\, & \max(\{s(x_1),s(x_3)\}) - s'(\{x_1,x_3\}) \leq \\
  & \Sep \max({\{s(x_1),s(x_2)\}}) - s'(\{x_1,x_2\}) \\
  & \Sep + \max({\{s(x_2),s(x_3)\}})- s'(\{x_2,x_3\}) \\
  \Rightarrow \,\,\, & \Db(x_1,x_3) \leq \Db(x_1,x_2) + \Db(x_2,x_3).
\end{align*}
\end{itemize}\SmallVJump

\newcommand{\Sp}[2]{s'\{#1, #2\}}
\newcommand{\Ms}[2]{\max\{#1, #2\}}
\renewcommand{\Sep}{\hspace{0.1cm}}
\item {(M3) \textsc{is valid for $\Dc$:}} \,\, The triangle inequality
  (M3) requires expression $\Dc(x_1,x_2) + \Dc(x_2, x_3) -
  \Dc(x_1,x_3)$ to be greater than or equal to zero. We split this
  into three cases:
  \begin{itemize}
  \item (Case 1) If $s(x_1) = s(x_2) = 0$, then, by
    Property~\ref{prop:uniqzero}, $x_1 = x_2$ and the triangle
    inequality becomes $\Dc(x_1,x_1) + \Dc(x_1, x_3) - \Dc(x_1,x_3)
    \geq 0$ which, by definition of $\Dc$, can be reduced to $\Dc(x_1,
    x_3) \geq \Dc(x_1,x_3)$ which is obviously true. An analogous
    argument can be made to show that (M3) is valid in the cases where
    $s(x_1) = s(x_3) = 0$ and $s(x_2) = s(x_3) = 0$.
  \end{itemize}
  If (Case 1) doesn't occur, then at least two elements in $\{x_1,
  x_2, x_3\}$ have size $s$ greater than zero and the definition of
  $\Dc$ we need to use is the bottom one in Equation~(\ref{eq:dc}). 
  In this case, the expression for (M3) becomes \hbox{(\ref{eq:dcm3})
    $\geq 0$}, where (\ref{eq:dcm3}) is
  \begin{align}
    \label{eq:dcm3}
    1  + \frac{\Sp{x_1}{x_3}}{M_{13}} - \frac{\Sp{x_1}{x_2}}{M_{12}} - \frac{\Sp{x_2}{x_3}}{M_{23}}
  \end{align}
  and $M_{ij}$ is a short name for $\Ms{s(x_i)}{s(x_j)}$.
  The remaining cases that are sufficient to prove that (M3) is valid
  for $\Dc$ are:
  \begin{itemize}
  \item (Case 2) If not (Case 1) and $s(x_2) \geq s(x_1), s(x_3)$ then
\begin{align*}
   & s(x_2) \times (\ref{eq:dcm3}) \\
   = \,\,\, & s(x_2) \times \Bigg(1  + \frac{\Sp{x_1}{x_3}}{M_{13}} - \frac{\Sp{x_1}{x_2}}{M_{12}} 
   - \frac{\Sp{x_2}{x_3}}{M_{23}}\Bigg)\\
    \geq  \,\,\, & s(x_2) + \Sp{x_1}{x_3}  - \Sp{x_1}{x_2} - \Sp{x_2}{x_3}\\
  \geq \,\,\, & 0, \text{by Lemma \ref{lem:aux}}.
\end{align*} Since $s(x_2) > 0$, this implies that $(\ref{eq:dcm3}) \geq 0$.
  \item (Case 3) Similarly, if not (Case 1) and $s(x_1) \geq s(x_2), s(x_3)$ then
\begin{align*}
            & s(x_1) \times (\ref{eq:dcm3}) \\
   = \,\,\, & s(x_1) \times \Bigg(1  + \frac{\Sp{x_1}{x_3}}{M_{13}} - \frac{\Sp{x_1}{x_2}}{M_{12}} 
                                                                   - \frac{\Sp{x_2}{x_3}}{M_{23}} \Bigg)\\
   = \,\,\, & s(x_1) \Bigg(1 - \frac{\Sp{x_2}{x_3}}{M_{23}}\Bigg) + \Sp{x_1}{x_3} - \Sp{x_1}{x_2} \\
   \geq \,\,\, & s(x_2) \Bigg(1 - \frac{\Sp{x_2}{x_3}}{M_{23}}\Bigg) + \Sp{x_1}{x_3} - \Sp{x_1}{x_2} \\
   = \,\,\, & s(x_2) - \frac{s(x_2)\Sp{x_2}{x_3}}{M_{23}} + \Sp{x_1}{x_3} - \Sp{x_1}{x_2} \\
   \geq \,\,\, & s(x_2) - \Sp{x_2}{x_3} + \Sp{x_1}{x_3} - \Sp{x_1}{x_2} \\
   \geq \,\,\, & 0, \text{by Lemma \ref{lem:aux}}.
\end{align*} Since $s(x_1) > 0$, this implies that $(\ref{eq:dcm3}) \geq 0$.
  \end{itemize}
      
\renewcommand{\Sep}{\hspace{0.1cm}}
\item {(M3) \textsc{is valid for $\Dd$:}} \,\, 
  This proof follow the same lines as the one given for
  graphs in \cite{wallis2001graph}. The triangle inequality
  (M3) requires expression $\Dd(x_1,x_2) + \Dd(x_2, x_3) -
  \Dd(x_1,x_3)$ to be greater than or equal to zero. We split this
  into three cases:
  \begin{itemize}
  \item (Case 1) If $s(x_1) = s(x_2) = 0$, then, by
    Property~\ref{prop:uniqzero}, $x_1 = x_2$ and the triangle
    inequality becomes $\Dd(x_1,x_1) + \Dd(x_1, x_3) - \Dd(x_1,x_3)
    \geq 0$ which can be reduced to \hbox{$\Dd(x_1,x_3) \geq
      \Dd(x_1,x_3)$} which is obviously true. An analogous argument
    can be made to show that (M3) is valid in the cases where $s(x_1)
    = s(x_3) = 0$ and $s(x_2) = s(x_3) = 0$.
  \end{itemize}
  If (Case 1) doesn't occur, then at least two elements in $\{x_1,
  x_2, x_3\}$ have size $s$ greater than zero and the definition of
  $\Dd$ we need to use is the bottom one in Equation~(\ref{eq:dc}). 
  In this case, the expression for (M3) becomes \hbox{(\ref{eq:ddm3})
    $\geq 0$}, where (\ref{eq:ddm3}) is
  \begin{align}
    \label{eq:ddm3}
    1  + \frac{\Sp{x_1}{x_3}}{U_{13}} - 
         \frac{\Sp{x_1}{x_2}}{U_{12}} - 
         \frac{\Sp{x_2}{x_3}}{U_{23}}
  \end{align}
  and $U_{ij} = s(x_i) + s(x_j) - \Sp{x_i}{x_j}$.
  Let $x_{ij} \in mcs(\{x_i,x_j\})$. By (A2) of a \MCSext\ there
  exists $x_{123} \OrdRel x_{12}, x_{23}$ such that $s(x_{123}) \leq
  s(x_{13}) = \Sp{x_1}{x_3}$ and $s(x_{123}) \geq s(x_{12}) + s(x_{23}) - s(x_2)$.
  In this way we can write
  \begin{align}
    (\ref{eq:ddm3})  = \,\,\, &  1  + \frac{s(x_{13})}{U_{13}} - \frac{s(x_{12})}{U_{12}} - \frac{s(x_{23})}{U_{23}} \nonumber \\
    \label{eq:ddm3aux}
    \geq \,\,\, &  1  + \frac{s(x_{123})}{s(x_1) + s(x_3) - s(x_{123})} - \frac{s(x_{12})}{U_{12}} - \frac{s(x_{23})}{U_{23}}.
  \end{align} 
  Let non negative numbers $a_1, a_2, a_3, a_{12}, a_{23}, a_{123}$ be
  defined by $s(x_{123}) = a_{123}$; $s(x_{12}) = a_{12} +
  a_{123}$; $s(x_{23}) = a_{23} + a_{123}$; 
  $s(x_1) = a_1 + a_{12} + a_{123}$; $s(x_2) = a_2 + a_{12}
  + a_{23} + a_{123}$; $s(x_3) = a_3 + a_{23} + a_{123};$. And let
  $T = a_1 + a_2 + a_3 + a_{12} + a_{23} + a_{123}$. We now can write
  \begin{align}
    \label{eq:ddm3aux2}
    (\ref{eq:ddm3aux}) = \,\,\, &  1  + \frac{a_{123}}{T-a_2} - \frac{a_{12} + a_{123}}{T-a_3} - \frac{a_{23}+ a_{123}}{T-a_1} 
  \end{align} 
  To show that $(\ref{eq:ddm3aux2}) \geq 0$ it is sufficient to show
  that $(\ref{eq:ddm3aux2})$ times a positive number is greater than
  or equal to zero. Let $(T-a_1)(T-a_2)(T-a_3)$ be this positive
  number, since (Case 1) is false. 
  \begin{align}
    & (\ref{eq:ddm3aux2}) \times (T-a_1)(T-a_2)(T-a_3) \nonumber \\
    \label{eq:ddm3aux3}
    = \,\,\, &  (T-a_1)(T-a_2)(T-a_3)  + a_{123}(T-a_1)(T-a_3) \nonumber \\
    & - (a_{12} + a_{123})(T-a_1)(T-a_2) \nonumber \\
    & - (a_{23} + a_{123})(T-a_2)(T-a_3) \nonumber \\
    = \,\,\, & a_1 a_2 (T - a_3) +  T \, (a_1 a_3 + a_2 a_3 + a_1 a_{12} + a_2 a_{12} \nonumber \\
    & \hspace{0.55cm}  + a_2 a_{123} + a_2 a_{23} + a_3 a_{23} + a_2 a_{123}  ) \nonumber \\
    & +  \, (a_1 a_3 a_{123} + a_1 a_2 a_{12} + a_1 a_2 a_{123} + a_2 a_3 a_{23} + a_2 a_3 a_{123} ) \nonumber \\
    \geq \,\,\, & 0 \nonumber
  \end{align}
\end{itemize} 
The proof of Theorem~\ref{thm:metrics} is complete. 
\end{Proof}

\section{Graph MCS Models}
\label{sec:graphmcs}

\newcommand{\Sgve}{s_{GVE\alpha}}
\newcommand{\Sgvep}{s'_{GVE\alpha}}
\newcommand{\Sgv}{s_{GV\alpha}}
\newcommand{\Sgvp}{s'_{GV\alpha}}
\newcommand{\Sges}{s_{GES}}
\newcommand{\Sgesp}{s'_{GES}}

In this section we present three examples of \MCSMS\ on graphs and
use Theorem~\ref{thm:metrics} to derive different metrics on graphs for
each of these examples. In particular, we are able to reproduce and
generalize previous metrics on graphs based on subgraphs and induced
subgraphs, and obtain new metrics on graphs based on an {\it extended}
subgraph notion.

\subsection{Graphs Terminology}

Here is a series of graph related definitions we use in the rest of
the paper. We chose undirected simple graphs as our default case, but
the results we present in the following sections also work for directed
graphs.

\Def{{Graph}}\label{def:graph} A {\it graph} is a 4-tuple $g=(V,E,\ell_v,\ell_e)$ 
  where
\vspace{-0.3cm}
\begin{itemize}
  \item $V$ is a finite set of {\it vertices};
  \item $E \subseteq [V]^2$ is the set of {\it edges};
  \item $\Lv:V \rightarrow \Sigma_V$ is a function that assigns labels to vertices;
  \item $\Le:E \rightarrow \Sigma_E$ is a function that assigns labels to edges;
\end{itemize}
If $V = \emptyset$ then $g$ is called the {\it empty graph}.
\EndDef

\Def{{Subgraph}} 
\label{def:subgraph}
A graph $g'=(V',E',\Lv',\Le')$ is said to be a {\it subgraph}
  of $g=(V,E,\Lv,\Le)$, if \, $V' \subseteq V$, \, $E' \subseteq E \cap [V']^2$,
  \, $\Lv'(v) = \Lv(v)$ for $v \in V'$,\, and $\Le'(e) = \Le(e)$ for $e \in E'$.
\EndDef

\Def{{Induced Subgraph}} A graph $g'=(V',E',$ $\Lv',\Le')$ is said to be
  an {\it induced subgraph} of $g=(V,E,\Lv,\Le)$ if $g'$ is a subgraph of $g$
  and $E'= E \cap [V']^2$.
\EndDef

\Def{{Isomorphism}} Let $g_1=(V_1,E_1,\Lv{_1},\Le{_1})$ and
$g_2 = (V_2, E_2,$ $\Lv{_2}, \Le{_2})$ be graphs. A bijection $\phi:V_1
\rightarrow V_2$ is an {\it isomorphism between $g_1$ and $g_2$}\, if
the conditions four conditions are valid: (1) $E_2 = \{ \{\phi(u),
\phi(v)\}\, : \, \{u,v\} \in E_1\}$; (2) $\Lv{_1}(v) =
\Lv{_2}(\phi(v)), \, \text{for } v \in V_1$; (3) $\Le{_1}(\{u,v\})$ $=$
$\Le{_2}(\{\phi(u),\phi(v)\}),$  for $\{u, v\} \in E_1$.  If
there exists an isomorphism between two graphs we say they are {\it
  isomorphic}.  \EndDef

\begin{rem} We use the notion $\phi (e)$, where $e= \{u,v\} \in E_1$ and $u,v\in V_1$,
to mean the edge $\{\phi(u),\phi(v)\} \in E_2$.
\end{rem}

\Def{{Subgraph Isomorphic}} \label{def:subgraphisomorphic}
  A graph $g$ is {\it subgraph isomorphic} to a graph $g'$, denoted by
  $g' \Subg g$, if there exists a subgraph of $g$ that is
  isomorphic to $g'$.
\EndDef

\Def{{Induced Subgraph Isomorphic}} \label{def:inducedsubgraphisomorphic}
  A graph $g$ is {\it induced subgraph isomorphic} to a graph $g'$,
  denoted by $g' \ISubgraph g$, if there exists an induced subgraph
  of $g$ that is isomorphic to $g'$.
\EndDef

\Def{{Graph $n$-Completion}} Let $g = (V,E,\ell_V,\ell_E)$ be a
graph with vertex labels in $\Sigma_{V}$ and edge labels in
$\Sigma_{E}$.  For $n \geq |V|$, a special vertex label $\Ev$,
and a special edge label $\Ee$, we define the {\it graph
  $n$-completion of $g$} as 
$$\Cn^{\Ev,\Ee}(g) = (V', E', \ell'_V,
\ell'_E)$$ where
\begin{itemize}
\item $V' = V \cup \{v_1, \ldots, v_{n-|V|}\}$,
\item $E' = [V']^2$,
\item $\ell'(v) = 
    \begin{cases} 
        \ell(v), &  \text{if }v \in V, \\
        \Ev& \text{if }v \in V'\backslash V, \\
    \end{cases}$
\item $\ell'(e) = 
    \begin{cases} 
        \ell(e), &  \text{if }e \in E, \\
        \Ee& \text{if }e \in E'\backslash E. \\
    \end{cases}$
\end{itemize}
When $\Ev$ and $\Ee$ are clear in the context, we will denote the
graph $n$-completion of $g$ as $\Cn(g)$.  
\EndDef

\subsection{Subgraph MCS Model}
\label{sec:gsmcs}
\vspace{0.2cm}

The first example of MCS Model on graphs is based on the subgraph
relation $\Subg$ (Definition~\ref{def:subgraphisomorphic}).

\Def{{S-MCS Model}} A {\it subgraph \MCSM}
or {\it \hbox{S-MCS} Model} is a triple
  $$(G,\Subgraph,\Sgve),$$ where 
\vspace{-0.2cm}
  \begin{itemize}
    \item $G$ is the set of graphs (Definition~\ref{def:graph}) with
      vertex labels in $\Sigma_V$, and edge labels in
      $\Sigma_E$. Furthermore, we consider two graphs $g_1, g_2 \in G$
      that are isomorphic to be the same graph: $g_1 = g_2$.
    \item $\Subg$ is the subgraph isomorphic relation on $G$
      (Definition~\ref{def:subgraphisomorphic});
    \item $\Sgve:G\rightarrow [0,+\infty)$ is a function
      based on a {\it label weighting function} $\alpha:(\Sigma_V \cup
      \Sigma_E)\rightarrow (0,+\infty)$ and, for $g=(V,E,\Lv,\Le)$, is
      defined by
      \begin{equation}
        \Sgve(g) = \begin{cases}
        0, \, \text{ if } V = \emptyset; \\[0.2cm]
        \displaystyle \sum_{v\in V}{\alpha(\Lv(v))} + \sum_{e\in E}{\alpha(\Le(e))}, \text{ otherwise.}
        \end{cases}
      \end{equation}
    \end{itemize}
\EndDef

The following theorem shows a S-MCS Model is indeed a MCS Model.

\begin{thm}
\label{thm:gsmom} The S-MCS Model is a MCS Model.
\end{thm}
\begin{Proof}
It can be verified that $\Subg$ is a partial order on $G$. Here
we are only going to show (S1), (S2), (A1) and (A2).
\begin{itemize}
\item[(S1)] Let $g_1 = (V_1, E_1, \Lvp{1}, \Lep{1})$ and $g_2 = (V_2, E_2,
  \Lvp{2}, \Lep{2})$ be graphs in $G$. If $g_1 \Subg g_2$ then there is an
  isomorphism $\phi$ between $g_1$ and $g_2'$, a subgraph of
  $g_2$. It should be clear that $s(g_1) = s(g'_2)$ since for every
  vertex and edge in $g_1$ there is a $\phi$-corresponding, equally labeled,
  vertex and edge in $g'_2$ and vice-versa. As the vertices and edges
  of $g'_2$ are subsets of $V_2$ and $E_2$, then $s(g'_2) \leq s(g_2)$, 
  since the vertex and edge sums in $\Sgve$ would run over these subsets.
  From this we can conclude $\Sgve(g_1) \leq \Sgve(g_2)$. 
\item[(S2)] Consider the same setup as in (S1) above: $g_1 \Subg g_2$
  and $g_1$ isomorphic to subgraph $g'_2$ of $g_2$. Add the extra
  hypothesis that $\Sgve(s_1) = \Sgve(s_2)$. This implies, as
  $\Sgve(g_1) = \Sgve(g'_2)$, that $\Sgve(g'_2) = \Sgve(g_2)$ which
  implies that the vertices and edges of $g'_2$ are exactily $V_2$ and
  $E_2$. In other words, $g'_2 = g_2$ and $g_1$ is isomorphic to $g_2$
  which in our case is the same as $g_1 = g_2$.
\item[(A1)] We are going to show that (A1') holds. In fact, the empty 
  graph is a subgraph of any other graph. This
  implies that, for any pair $g_1, g_2 \in G$, we have
  $\{\text{``empty graph''}\} \subseteq \CSubpp{g_1,g_2}$ and,
  consequently, $0 < 1 \leq |\CSubpp{g_1,g_2}|$. Also, by our
  definition, graphs have finite number of vertices and edges. This
  implies that all subgraphs of a graph is also finite (all possible
  subsets of the vertex and edge sets of a graph are finite). As we
  are consider isomorphic graphs to be equal, we know $\CSubpp{g_1,
    g_2} \subseteq \CSubpp{g_1} = \text{``subgraphs of $g1$''}$. As
  the right set in the previous chain is finite implies $\CSubpp{g_1,
  g_2}$ is also finite.
\item[(A2)] Let $g_1, g_2 \Subg h$. Let $\phi_1$ be an isomorphim
  between $g_1$ and a subgraph
  $h_1=(V_{h_1},E_{h_1},\Lvp{h_1},\Lep{h_1})$ of $h$ and $\phi_2$ be
  an isomorphism between $g_2$ and a subgraph
  $h_2=(V_{h_2},E_{h_2},\Lvp{h_2},\Lep{h_2})$ of $h$. Define $h_{12}$
  to be another subgraph of $h$ whose vertices and edges are,
  respectively, $V_{h_1} \cap V_{h_2}$ and $E_{h_1} \cap E_{h_2}$.
  The vertex and edges labels of $h_{12}$ are chosen to match
  the ones in $h$. With this construction of $h_{12}$ it can be
  verified that $h_{12} \Subg g_1, g_2$ and that 
  $$\Sgve(h) \geq \Sgve(g_1) + \Sgve(g_2) - \Sgve(h_{12}).$$
  To see this last inequation one should only notice that every
  vertex and edge of $h$ that was counted twice in the sum
  $\Sgve(g_1) + \Sgve(g_2)$ is decreased once when we subtract  
  $\Sgve(h_{12})$.
\end{itemize}
This completes the proof of Theorem~\ref{thm:gsmom}.
\end{Proof}

The Theorem \ref{thm:gsmom} is true, if we use directed graph instead
of undirected ones. Hence, we conclude, as a corollary of Theorem
\ref{thm:gsmom}, that \hbox{$(G,\Subgraph,s_{GVE_{\alpha_1}})$},
where $\alpha_1$ denotes the constant function equal to one, is a MCS
model and hence $\DFV$ is a metric on $G$ (since, by Theorem
\ref{thm:metrics}, $\Da$ is a metric on $G$), with this we reobtain
the result in \cite{fernandez2001graph}.  Furthermore, again by
Theorem \ref{thm:metrics}, $\Dc$ and $\Dd$ are metrics on $G$, which
shows versions of $\DBS$ and $\DW$ based on subgraphs (note that we
need to use as size function the sum of the number of edges and the
number of vertices).

It is worth noting that the Theorem \ref{thm:gsmom} enables the use of
different label weighting functions that makes possible to enconde
application domain knowledge in the \MCSM\ definition and hence in
the metrics in Theorem~\ref{thm:metrics}.

\subsection{Induced Subgraph MCS Model}
\label{sec:gismcs}

The second example of MCS Model on graphs is based on the induced
subgraph relation $\ISubgraph$ (Definition~\ref{def:inducedsubgraphisomorphic}).

\Def{{I-MCS Model}}An {\it induced subgraph \MCSM}
or {\it \hbox{I-MCS} Model} is a triple $$(G,\ISubgraph,\Sgv),$$ where 
\vspace{-0.2cm}
  \begin{itemize}
    \item $G$ is the set graphs (Definition~\ref{def:graph}) with
      vertex labels in $\Sigma_V$, and edge labels in $\Sigma_E$.
      Furthermore, we consider two graphs $g_1, g_2 \in G$
      that are isomorphic to be the same graph: $g_1 = g_2$.
    \item $\ISubgraph$ is the induced subgraph relation on $G$
      (Definition~\ref{def:inducedsubgraphisomorphic});
    \item and $\Sgv:G\rightarrow [0,+\infty)$ is a function
      based on a {\it label weighting function} $\alpha:\Sigma_V 
      \rightarrow (0,+\infty)$ and, for $g=(V,E,\Lv,\Le)$, is
      defined by
      \begin{equation}
        \Sgv(g) = \begin{cases}
        0, \, \text{ if } V = \emptyset; \\[0.2cm]
        \displaystyle \sum_{v\in V}{\alpha(\Lv(v))}, \text{ otherwise.}
        \end{cases}
      \end{equation}
    \end{itemize}
\EndDef


The following theorem shows that I-\MCSMS\ are indeed a MCS Model.

\begin{thm}\label{thm:graphsubgraphSOS} The I-MCS Model is a MCS Model.
\end{thm}
\begin{Proof} 
It can be verified that $\ISubg$ is a partial order on $G$. The
arguments to show that (S1), (S2) and (A1) are valid here are
essentially the same as in the proof of Theorem~\ref{thm:gsmom} if we
replace the terms ``subgraph'' by ``induced subgraph'' and $\Sgve$ by
$\Sgv$.  For (A2) it is sufficient to notice that the construction of
$h_{12}$ in the other proof replacing ``subgraph'' by ``induced
subgraph'' yields an induced subgraph of $g_1$ and $g_2$ and the same
argument used there to show the (A2) inequation was valid with
$h_{12}$ can also be used here.  
\end{Proof}

Again, Theorem \ref{thm:graphsubgraphSOS} also holds if we deal with
directed graphs. Thus, we can get as corollary of Theorem
\ref{thm:gsmom} the fact that $(G,\ISubgraph, s_{GV_{\alpha_1}})$ is a
\MCSM, where $\alpha_1$ denotes the constant function equal to
one. Thus, $\DBS$ and $\DW$ are metrics on $G$ (since, by Theorem
\ref{thm:metrics}, $\Dc$ and $\Dd$ are metrics on $G$), with this we
reobtain the results by \cite{bunke1998gdm} and
\cite{wallis2001graph}. Furthermore, since $\Da$ is a metric on $G$,
we get a version of $\DFV$ based on the induced subgraph relation
(using as size function the number of vertices).  Also, as in the
previous case, the use of different label weighting function allows
application domain knowledge to be used in the definition of the
metrics (similarity notion).

\FigureTwo

In Figure~\ref{fig:diff-mcs-notions} we illustrate the effect on the
notion of maximum common subelement (subgraphs and induced subgraphs)
caused by using different label weighting functions $\alpha$.

\subsection{{Extended Subgraph MCS Model}}
\label{sec:gesmcs}
\vspace{0.2cm}

The previous two examples of \MCSMS\ on graphs were based on two
well known partial orders on graphs: subgraphs and induced subgraphs.
We now define a third kind of partial order on graphs that we call
{\it extended subgraphs}. The idea is a simple generalization of the
subgraph partial order. Suppose that we fix partial orders on the
vertex and edge label sets of a graph. Informally we say that a graph
$g_1$ is an extended subgraph of a graph $g_2$ with respect to these
label partial orders, if we can fit the structure of $g1$ into $g_2$
in a way that each aligned vertex and edge has a label in $g_1$ that
is a subelement (by the label partial order) of the corresponding
aligned element in $g_2$. This informal idea is defined preciselly in
the following two definitions.

\Def{{Extended Subgraph}}\label{def:esubgraph} Let
$g'=(V',E',\Lv',\Le')$ and $g=(V,E,\Lv,\Le)$ be graphs with vertex
labels in $\Sigma_V$ and edge labels in $\Sigma_E$. If
$\OrdRel_{\Sigma_V}$ is a parital order on $\Sigma_V$ and
$\OrdRel_{\Sigma_E}$ is a partial order on $\Sigma_E$, we
say {\it $g'$ is an extended subgraph of $g$ with respect to
  $\OrdRel_{\Sigma_V}$ and $\OrdRel_{\Sigma_E}$} if 
\begin{align*}
  & V' \subseteq V, \,\, E' \subseteq E \cap [V']^2, \\
  & \Lv'(v) \OrdRel_{\Sigma_V} \Lv(v), \text{ for } v \in V', \\
  & \Le'(e) \OrdRel_{\Sigma_E} \Le(e), \text{ for } e \in E'.
\end{align*}
When $\OrdRel_{\Sigma_V}$ and $\OrdRel_{\Sigma_E}$ are clear in the
context we simply say that {\it $g'$ is an extended subgraph of $g$}.
\EndDef

\Def{{Extended Subgraph Isomorphic}} \label{def:esubgraphiso} If $g'$ is
isomorphic to a graph that is an extended subgraph of $g$ with respect
to $\OrdRel_{\Sigma_V}$ and $\OrdRel_{\Sigma_E}$, we say that $g$ is
{\it extended subgraph isomorphic} to $g'$, and denote this fact
by $g' \ESubgraph g$.
\EndDef

We are now able to define the third example of MCS Model on graphs
based on the extended subgraph relation $\ESubgraph$
(Definition~\ref{def:esubgraphiso}).




\Def{{E-MCS Model}} Let $\MV$ and $\ME$ be \MCSMS
on $\SV$ and $\SE$
\begin{align*}
  & \MV = \MVMOM, \\
  & \ME = \MEMOM,
\end{align*}
with size functions being strictly positive: $\SizeSV>0$ and
$\SizeSE>0$. An extended subgraph MCS or E-MCS Model with respect to
$\MV$ and $\ME$ is a triple
$$(G, \ESubgraph, \Sges)$$
where,
\vspace{-0.2cm}
  \begin{itemize}
    \item $G$ is the set of graphs (Definition~\ref{def:graph}) with
      vertex labels in $\SV$, and edge labels in $\SE$; Furthermore,
      we consider two graphs $g_1, g_2 \in G$ that are isomorphic to
      be the same graph: $g_1 = g_2$.
    \item $\ESubgraph$ is the extended subgraph relation on $G$
      with respect to $\SubSV$ and $\SubSE$  (Definition~\ref{def:esubgraphiso});
    \item and for $g=(V,E,\Lv,\Le) \in G$,
      \begin{equation}
        \label{eq:sges}
      \Sges(g) = 
      \begin{cases}
        \,\, 0\,, \text{ if } V = \emptyset; \\[0.2cm]
        \,\, \displaystyle \sum_{v\in V}{s_{\Sigma_V}(\Lv(v))} + \sum_{e\in E}{s_{\Sigma_E}(\Le(e))}, \,\,  \text{ otherwise.}
      \end{cases}
      \end{equation}
  \end{itemize}
\EndDef


The following theorem shows that E-\MCSMS\ are indeed MCS Models.

\begin{thm}\label{thm:gesmom} 
  The E-MCS Model is a MCS Model.
\end{thm}
\begin{Proof} 
The proof goes as follows:
\begin{itemize}
\item[(S1)] Let $g_1, g_2$ be graphs such that $g_1 \OrdRelEG g_2$. By
  definition of $\OrdRelEG$, there exists an extended subgraph $g'_2$
  of $g_2$ that is isomorphic to $g_1$, and, clearly, $\Sges(g_1) =
  \Sges(g'_2)$. Since $\Sges(g'_2)$ is a sum running over a subset of
  the vertices and edges of $\Sges(g'_2)$, and each vertex or edge in
  the sum of $\Sges(g'_2)$ yields a smaller or equal value than the
  one in the sum of $\Sges(g_2)$, we can conclude that (S1) is valid.
\item[(S2)] Let $g_1 \OrdRelEG g_2$ and $s(g_1) = s(g_2)$.  By
  definition of $\OrdRelEG$, there exists an extended subgraph $g'_2$
  of $g_2$ that is isomorphic to $g_1$ and, clearly, $\Sges(g_1) =
  \Sges(g'_2)$. This implies $\Sges(g'_2) = \Sges(g_2)$. As $g'_2$ is
  a subgraph of $g_2$ the only option to make $\Sges(g'_2) =
  \Sges(g_2)$ is to have $g_2 = g'_2$.
\item[(A1)] Again, let $g_1 = (V_1,E_1,{\Lv}_1,{\Le}_1)$ and $g_2 =
  (V_2,E_2,{\Lv}_2,{\Le}_2)$ be two graphs.  The fact that the empty
  graph is a subgraph of any graph implies that $\{\text{empty
    graph}\} \subseteq \CSubpp{g_1,g_2}$ and, consequently, $0 <
  |\CSubpp{g_1,g_2}|$. In order to prove that the set $\{s(g)|g
  \OrdRelEG g_1,g_2\}$ has a maximum, let $g_1^\emptyset = (V_1,E_1)$
  and $g_2^\emptyset = (V_2,E_2)$ be the unlabelled copies (same
  structure) of $g_1$ and $g_2$ respectivelly. The set of common
  subgraphs (by the subgraph isomorphic relation) of $g_1^\emptyset$
  and $g_2^\emptyset$ is finite, as in the proof of Theorem
  \ref{thm:gsmom}.  Denote this set by $cs(g_1^\emptyset,
  g_2^\emptyset) = \{h_1,...,h_n\}$. For each $h_i$, let $\Phi_i =
  \{\phi_i¹,...,\phi_i^{k_i}\}$ be the set of all subgraph
  isomorphisms between $h_i$ and $g_1$. Similarly, let $\Psi_i =
  \{\psi_i¹,...,\psi_i^{l_i}\}$ be the set of all subgraph
  isomorphisms between $h_i$ and $g_2$. Now, for each $s=1,...,k_i$
  and $t=1,...,l_i$, the map $\phi_i^s \circ (\psi_i^t)^{-1}$ defines
  a isomorphism from a subgraph $g_1^{st}$ of $g_1^\emptyset$ and a
  subgraph $g_2^{st}$ of $g_2^\emptyset$. Finally, denote by $g_{st}$
  the extended subgraph of $g_1$ and $g_2$ that has the vertex and
  edge set the same as $g_1^{st}$ and the labels are defined as
  follows: for each vertex $v$ and edge $e$ of $g_{st}$ define its
  label as an element of $mcs({\Lv}_1(v), {\Lv}_2(\phi_i^s \circ
  (\psi_i^t)^{-1}(v)))$ and $mcs({\Le}_1(e), {\Le}_2(\phi_i^s \circ
  (\psi_i^t)^{-1}(e)))$, respectively. Now let $s_0$ and $t_0$ be such
  that $\Sges(g_{s_0t_0}) = max(\{\Sges(g_{st})|s=1,...,k_i \text{ and
  } t=1,...,l_i\})$.  By construction, such $g_{s_0t_0}$ is the
  extended subgraph of $g_1$ and $g_2$ with maximum size.
\item[(A2)] Let $g_1, g_2 \ESubgraph h$. Let $\phi_1$ be an isomorphim
  between $g_1$ and an extended subgraph
  $h_1=(V_{h_1},E_{h_1},\Lvp{h_1},\Lep{h_1})$ of $h$ and $\phi_2$ be
  an extended isomorphism between $g_2$ and a subgraph
  $h_2=(V_{h_2},E_{h_2},\Lvp{h_2},\Lep{h_2})$ of $h$. Define $h_{12}$
  to be another extended subgraph of $h$ whose vertices and edges are,
  respectively, $V_{h_1} \cap V_{h_2}$ and $E_{h_1} \cap E_{h_2}$.
  The label of vertex $v$ of $h_{12}$ is defined as follows: let
  $\sigma_{h}, \sigma_{h_1}, \sigma_{h_2}$ be the label of $v$ in,
  respectively, $h, h_1, h_2$; by the fact that $h_1, h_2$ are
  extended subgraphs of $h$, we have that $\sigma_{h_1}, \sigma_{h_2}
  \OrdRel_{\SV} \sigma_h$; by axiom (A2) in $\Model{V}$ there exist
  $\sigma_{h_{12}} \in \SV$ such that $s_{\SV}(\sigma_h) \geq
  s_{\SV}(\sigma_{h_1}) + s_{\SV}(\sigma_{h_2}) -
  s_{\SV}(\sigma_{h_{12}})$; define the label of $v$ in $h_{12}$ to be
  $\sigma_{12}$. The label of an edge of $h_{12}$ is defined
  in an analogous way. With this construction of $h_{12}$ it can be
  verified that $h_{12} \Subg g_1, g_2$ and that 
  $$\Sges(h) \geq \Sges(g_1) + \Sges(g_2) - \Sges(h_{12}).$$
\end{itemize}
The proof of Theorem~\ref{thm:gesmom} is complete.
\end{Proof}


When modeling real world concepts using graphs, it is usually
important to have flexibility when defining what information a vertex
or an edge will carry. For example, in scientific workflow
descriptions vertices represent parameterized modules that represent
some kind of computation. Usually a single module is configured with a
set of parameters and values which are not adequately represented by a
single symbol, but, instead, by a more complicated object. The {\it
  nesting property} of E-\MCSMS that enables plugging other MCS
Model elements as labels of vertices and edges, and be able to
derive metrics for these objects that take into account all parts that
form the final object is an interesting one. 

To illustrate E-MCS Models, we will use them, in next section, to
build a link between Graph Edit Distance and \MCSMS.  Before that
we need an additional property of E-\MCSMS\ that states that if we
restrict the elements (graphs) of an E-MCS Model to complete graphs of
$n$ vertices, we still have a MCS Model. We will refer to this MCS
model as a {\it $n$-restricted E-MCS Model}.

\newcommand{\OKN}{\OrdRel_{K_n}}
\newcommand{\SKN}{s_{K_n}}

\begin{prop}\label{prop:completemom} 
Let $\Model{}=(G, \ESubgraph, \Sges)$ be an E-MCS Model with respect
to $\MV=\MVMOM$ and $\ME=\MEMOM$. Let $K_n$ be the subset of $G$
formed of complete graphs with $n$ vertices. Let $\OKN$ and
$\SKN$ be the restrictions of $\ESubgraph$ and $\Sges$ to 
$K_n$. In this context, the triple \hbox{$\Model{K_n}=(K_n, \OKN,
\SKN)$} is also a MCS Model.
\end{prop}
\vspace{-0.2cm}
\begin{Proof} 
Properties (R1),(R2) and (R3) clearly hold for $\OKN$, since they are
valid for $\ESubgraph$. Similarly, the properties (S1) and (S2) hold
for $\SKN$, since they hold for $\Sges$.
\begin{itemize}
\item{(A1)} Let $g_1 = (V_1,E_1,{\ell_V}_1,{\ell_E}_1), g_2 =
  (V_2,E_2,{\ell_V}_2,{\ell_E}_2) \in K_n$.  Let $\phi$ be a bijection
  between $V_1$ and $V_2$. Then, $\phi$ defines a one-to-one
  correspondence between any vertex and edge of $g_1$ and $g_2$. Then,
  we can define a graph $g_{12} \in K_n$, by using the same vertex and
  edge sets as in $g_1$ and defining the label for each vertex $v \in
  V_1$ as an element of $cs(\{ {\ell_V}_1(v),{\ell_V}_2(\phi(v))\})$
  and for each edge $e \in E_1$ as an element of $cs(\{
  {\ell_E}_1(e),{\ell_E}_2(\phi(e))\})$.

  Now let $\{\phi_1,...,\phi_{n!}\}$ be the set of all bijection
  between $V_1$ and $V_2$.  For each bijection $\phi_k$, we can define
  a extended subgraph $g_{12}^k$ of $g_1$ and $g_2$ as before, but
  chosing as labels for each vertex an element of $mcs(\{
  {\ell_V}_1(v),$ ${\ell_V}_2(\phi_k(v))\})$ and for each edge
  an element of $cs(\{ {\ell_E}_1(e),$ ${\ell_E}_2(\phi_k(e))\})$.  Let
  $k_0$ be the index associated with the largest graph among
  $g_{12}^k$, {\it i.e.}, $s(g_{12}^{k_0}) = max(\{s_{K_n}(g_{12}^k) |
  k=1,...,n!)\})$. By construction, $s(g_{12}^{k_0}) =$ \linebreak
  $max(\{s_{K_n}(g) | g \in cs(\{g_1,g_2\}) \})$.

\item{(A2)} Let $g_1 = (V_1,E_1,{\Lv}_1,{\Le}_1), g_2 = (V_2,E_2,
  {\Lv}_2,{\Le}_1) \in K_n$ and let also $g = (V,E, \Lv,\Le) \in K_n$
  be such that $g_1,g_2 \OrdRel_{K_n} g$.  We want to define a
  complete graph $g_{12} = (V_{12},E_{12}, \Lv{_{12}},\Le{_{12}})$
  such that $g_{12} \OrdRel_{K_n} g_1,g_2 \OrdRel_{K_n} g$ and also that
  $s(g) \geq s(g_1) + s(g_2) - s(g_{12})$. In order to do so, we fix
  one correspondences $\phi_{12}$ (bijection) between $V_{12}$ and
  $V_1$ and other one $\phi_1$ between $V_1$ and $V_2$.  We are going
  to denote $v \in V_{12}, \phi_{12}(v)$ and $\phi_1(\phi_{12}(v))$,
  just by $v$. We define the label of $g_{12}$ as follows: For each $v
  \in V_{12}$ we know that ${\Lv}_1(v),{\Lv}_2(v) \OrdRel \Lv(v)$,
  then we can use the axiom (A2) for the \MCSM
  $(\Sigma_V,\OrdRel_{\Sigma_V},s_{\Sigma_V})$ and conclude that there
  exists a label $\alpha_{12}$ such that
  $\alpha_{12} \OrdRel_{\Sigma_V} \Lv{_1}(v),\Lv{_2}(v)
  \OrdRel_{\Sigma_V} \Lv(v)$ and
  $s_{\Sigma_V}(\Lv(v)) \geq s_{\Sigma_V}(\Lv{_1}(v)) +
  s_{\Sigma_V}(\Lv{_2}(v)) - s_{\Sigma_V}(\alpha_{12})$.  We define
  $\Lv{_{12}}(v) = \alpha_{12}$. With a similar construction, we
  can define the edge label function $\Le{_{12}}$. One can verify that 
  $g_{12}$ constructed this way satisfy the axiom (A2).
\end{itemize}
\end{Proof}




\section{Relation between Graph Edit Distance and MCS Models}
\label{sec:ged}

In this section we show a relation between graph edit distance and MCS
Models. Informally speaking, this connection states that if $\Dged$ is
a graph edit distance and a metric on $G$, then there is a
corresponding \MCSM\ $\Model{}$ such that $$\Dged(g_1,g_2) =
\Da(\theta(g_1),\theta(g_2)),$$ where $g_1, g_2 \in G$, $\theta$ takes
the elements of $G$ into their corresponding elements in $\Model{}$,
and $\Da$ is the first of the four metrics in
Theorem~\ref{thm:metrics} valid in $\Model{}$. Thus, the MCS Model
$\Model{}$ encodes $\Dged$. The problem of finding the graph edit
distance between $g_1$ and $g_2$ becomes the problem of finiding a
maximum common subelement between $\theta(g_1)$ and $\theta(g_2)$ in
$\Model{}$.

Before stating the main result of this section, we define precisely
what we mean by graph edit distance. We use the notion of graph
completion in this definition to facilitate the exposition: for any
two graphs we can refer to bijections between the vertices of their
completed versions instead of having to deal with functions between
subsets of the vertices of the first graph into the vertices of the
second graph. This definition of graph edit distance is equivalent to
the common use of the term, where the cost of each vertex and edge
operation (\ie addition, deletion, and substitution) is based on
labels and these operation costs are known a priori.


\Def{{Graph Edit Distance}} \,\, Let $g_1$ and $g_2$ be graphs with vertex labels in
$\Sigma_V$ and edge labels in $\Sigma_E$. Furthermore, let
\begin{align*}
& c_V: (\Sigma_V \cup \{\Ev\})^2 \rightarrow [0,\infty],  \\
& c_E: (\Sigma_E \cup \{\Ee\})^2 \rightarrow [0,\infty]
\end{align*}
be, respectivelly, {\it edit cost functions} on vertex and edge
labels, where $\Ev$ and $\Ee$ are special labels. 
Assume that
\begin{align*}
& g'_1 = \GC{|V_1| + |V_2|}{\Ev,\Ee}(g_1) = (V'_1,E'_1,\ell'_{V1},\ell'_{E1}), \\
& g'_2 = \GC{|V_1| + |V_2|}{\Ev,\Ee}(g_2) = (V'_2,E'_2,\ell'_{V2},\ell'_{E2}).
\end{align*}
Let ${\cal F}$ be the set of bijections from $V'_1$ to $V'_2$. 
The cost $c(f)$ for $f \in {\cal F}$ is defined as
\begin{equation}
\label{eqn:gedcost}
c(f) =   \sum_{v \in V'_1}{ c_V ( \ell'_{V1}(v), \ell'_{V2}( f(v) ) ) } \,\, + \,\, 
        \sum_{e \in E'_1}{ c_E ( \ell'_{V1}(e), \ell'_{E2}( f(e) ) ) }.
\end{equation}
In this context, we define the {\it graph edit distance} between $g_1$
and $g_2$ as
$$\Dged(g_1,g_2) = \min_{f \in {\cal F}} c(f).$$
\EndDef

Some uses of the term graph edit distance refer to a more general
idea. For example, \cite{bunke1997relation} shows a correspondence
between the maximum number of vertices of a common induced subgraph
and a specific graph edit distance notion where the edge operation
cost depends on which operation was done in its end vertices.

Now we are able to state the main result of this section.

\begin{thm}[{\sc GED and MCS Model}]
\label{thm:ged2rg1}
Let $G_n$ be the set of graphs with $n$ or less vertices on finite
label sets $\SV$ and $\SE$.  Let $c_V:(\SV \cup \Ev)^2 \rightarrow
[0,\infty)$ and $c_E:(\Sigma_E \cup \Ee)^2 \rightarrow [0,\infty)$ be
    edit cost functions. Furthermore, let $c_V$ and $c_E$ be metrics
    on $\SV \cup \{\Ev\}$ and $\SE \cup \{\Ee\}$.  Then, there exists
    a \MCSM\ $${{\cal{M}}_n} = (X,\Ox,s_X)$$ and an injective function
    $\theta: G_n \rightarrow X$ such that
\begin{align*}
\Dged(g_1, g_2) = s_X(\theta(g_1)) + s_X(\theta(g_2)) - 2 s'_X(\{\theta(g_1),\theta(g_2)\}).
\end{align*}
\end{thm}
\begin{Proof}
Apply Lemma~\ref{lem:dm2mom} on the finite metric spaces
\hbox{$(\SV\cup\{\Ev\}, c_V)$} and $(\SE\cup\{\Ee\}, c_E)$ to
obtain corresponding \MCSMS\ $\MV = (\SV',\OrdRel_{\SV'},s_{\SV'})$
and $\ME = (\SE',\OrdRel_{\SE'},s_{\SE'})$, where the size of the
smallest element in these \MCSMS\ are strictly positive. Let $G'$
be the set of graphs with labels in $\SV'$ and $\SE'$.  Observe that
the triple $(G', \ESubgraph, s_{GES})$ with respect to $\MV$ and $\ME$
is an E-MCS Model. Define $X=K_{2n}$ to be the subset of $G'$
consisting only of complete graphs with $2n$ vertices. By
Proposition~\ref{prop:completemom} we know that
$\Model{n}=(X,\OrdRel_X,s_X)$ is a MCS Model, when $\OrdRel_X$ and
$s_X$ are restrictions of $\ESubgraph$ and $s_{GES}$ to the set $X$.
Assume $g_1, g_2 \in G_n$ and their vertex sets are, respectively,
$V_1$ and $V_2$. By definition, the graph edit distance between $g_1$
and $g_2$ is the minimum value of function $c$
(Equation~\ref{eqn:gedcost}) for a vertex bijection between $\GC{|V_1|
  + |V_2|}{\Ev,\Ee}(g_1)$ and $\GC{|V_1| + |V_2|}{\Ev,\Ee}(g_2)$. It
can be checked that for our metric $c_V$ and $c_E$ this minimum value
of function $c$ is the same if we consider vertex bijections between
$\GC{2n}{\Ev,\Ee}(g_1)$ and $\GC{2n}{\Ev,\Ee}(g_2)$. Define
$\theta:G_n\rightarrow X$ to be the graph completion $\Cnn^{\Ev,
  \Ee}$. Make \hbox{$x_1 = \theta(g_1) =
  (V_1',E_1',\ell_{V_1}',\ell_{E_1}')$} and \hbox{$x_2 = \theta(g_2) =
  (V_2',E_2',\ell_{V_2}',\ell_{E_2}')$}. Let $f$ be a bijection
between the vertices of $x_1 $ and $x_2$. Define $x_f \in X$ in the
following way: for every vertex $v$ in $x_1$ there corresponds a
vertex in $x_f$ labeled with an element (any element) of
$mcs(\{\ell_{V_1}'(v),\ell_{V_2}'(f(v))\})$, and for every edge $e$ in
$x_1$ there corresponds an edge in $x_f$ labeled with an element (any
element) of $mcs(\{\ell_{E_1}'(e),\ell_{E_2}'(f(e))\})$. Using this
construction for $x_f$ it is clear that $x_f \OrdRel_X x_1, x_2$ and
it can be verified that $c(f) = s_X(x_1) + s_X(x_2) - 2 s_X(x_f)$. Let
$f_0$ be the bijection between vertices of $x_1$ and $x_2$ that yields
the graph edit distance between $g_1$ and $g_2$. At this point we can
write
$$d_{GED}(g_1,g_2) = c(f_0) = s_X(x_1) + s_X(x_2) - 2 s_X(x_{f_0}).$$
To conclude the proof it remains showing that $s_X(x_{f_0}) =
s_X'(\{x_1,x_2\})$. Assume $s_X(x_{f_0}) < s_X'(\{x_1,x_2\})$ and
$x_{12} \in mcs(\{x_1,x_2\})$. Let $\phi_1$ and $\phi_2$ be
isomorphisms between $x_{12}$ and extended subgraphs of $x_1$ and
$x_2$. Define $f_{12} = \phi_2 \circ \phi_1^{-1}$. Note that $f_{12}$
is a bijection between vertices of $x_1$ and of $x_2$ and that
$s_X(x_{12}) = s_X(x_{f_{12}})$.  In this case, we can write
$s_X(x_{f_0}) < s_X'(\{x_1,x_2\}) = s_X(x_{12}) = s_X(x_{f_{12}})$,
for bijection $f_{12}$.  This contradicts the hypothesis that
$s_X(x_{f_{0}})$ is the maximum possible for a bijection between
vertices of $x_1$ and $x_2$.  The theorem is proven.
\end{Proof}

An interesting aspect of this theorem is that it brings a different
and precise materialization for the meaning of a metric graph edit
distance between two graphs: we can see it encoded in an element of a
corresponding MCS Model. We see as applications of this connection,
the interpretation of natural notions in the MCS Model in terms of the
original metric graph edit distance. For example, a maximum common
subelement of three or more elements of the MCS Model could correspond
to a natural generalization of the metric graph edit distance between
three or more graphs.

\section{Conclusions}

In this paper we have introduced \MCSM\ which is a generalization of a
model proposed by \cite{deraedt2009ddm}. We then showed four metric
functions to be valid in any MCS Model (three additional metrics to
the one shown in \cite{deraedt2009ddm}). The usefulness of the MCS
Model is that it serves as a template to fit into applied scenarios
and ease the derivation of metrics (precise similarity notions) in
those scenarios. We show this usage of MCS Models by presenting three
examples on graphs: the S-MCS (based on subgraphs), I-MCS (based on
induced subgraphs), and E-MCS (based on a less common partial order
that we name extended subgraphs). With these examples we are able to
reproduce and extend previous reported metrics on graphs
\cite{bunke1998gdm, wallis2001graph, fernandez2001graph} as well as
new ones (\eg subgraph versions of $\DBS$ and $\DW$). The E-MCS Model
has an interesting nesting property that allows one to derive distance
metric for graphs with complex labels, which might be of important
value when modeling real scenarios.  A final contribution of this
paper is an interpretation of the graph edit distance that is a metric
on graphs as, essentially, a maximum common subelement on a
corresponding MCS Model.


\bibliographystyle{model2-names}
\bibliography{paper}  
\end{document}